\documentclass[a4paper]{article}

\usepackage[utf8]{inputenc}
\usepackage{float}

\usepackage{amsmath}
\usepackage{amsthm}
\usepackage{amssymb}
\usepackage{booktabs}
\usepackage[hidelinks]{hyperref}
\usepackage[textwidth=16cm,textheight=23cm]{geometry}

\usepackage{colortbl}
\usepackage{tabularx}
\usepackage{array}
\usepackage{tikz}
\usepackage{pgfplots}
\usepackage{pgfplotstable}
\usepackage{subcaption}
\usepackage{algpseudocode}
\usepackage{algorithm}
\pgfplotsset{compat=1.15}

\usepackage{import}
\usetikzlibrary{calc}
\usetikzlibrary{arrows,shapes}
\usetikzlibrary{positioning}
\usetikzlibrary{trees}
\usetikzlibrary{backgrounds}
\usetikzlibrary{arrows,shapes}
\usetikzlibrary{mindmap}
\usetikzlibrary{chains}
\usetikzlibrary{calc}
\usetikzlibrary{matrix}
\usetikzlibrary{decorations.pathmorphing,patterns,shadows}
\usepgfplotslibrary{fillbetween}
\usepackage{pgfplots}
\usepackage{pgfplotstable}

\pgfplotsset{compat=1.17}

\newcommand{\average}[1]{\left\{\!\left\{#1\right\}\!\right\}}
\newcommand{\jump}[1]{\left[\!\left[{#1}\right]\!\right]}

\definecolor{lowPrec}{HTML}{BEE0FF}  
\definecolor{highPrec}{HTML}{FFD8BE} 

\makeatletter
\let\@fnsymbol\@arabic

\definecolor{gnuplot@orange}{RGB}{229,158,0}
\definecolor{gnuplot@purple}{RGB}{148,0,212}
\definecolor{gnuplot@lightblue}{RGB}{87,181,232}
\definecolor{gnuplot@green}{RGB}{0,158,115}
\definecolor{gnuplot@darkblue}{RGB}{0,115,179}
\definecolor{gnuplot@yellow}{RGB}{240,227,66}
\pgfplotscreateplotcyclelist{colorGPL}{%
	gnuplot@darkblue,every mark/.append style={fill=gnuplot@darkblue!80!black},mark=o\\%
	red!80!white,every mark/.append style={fill=red!80!black},mark=square\\%
	gnuplot@green,every mark/.append style={fill=gnuplot@green!50!black},mark=otimes*\\%
	gnuplot@orange,every mark/.append style={fill=gnuplot@orange!80!black,mark size=2.5pt},mark=x\\%
	gnuplot@purple,mark=diamond\\%
	black,densely dashed,every mark/.append style={solid,fill=gnuplot@darkblue!80!black},mark=square*\\%
	gnuplot@lightblue,densely dashed,every mark/.append style={solid,fill=gnuplot@lightblue!30!black},mark=triangle*\\%
	red!80!white,densely dashed,every mark/.append style={solid,fill=red!40!white},mark=oplus*\\%
}
\makeatother

\newcommand{\AlgStep}[3]{%
  \rowcolor{#1}%
  \textbf{#2.}&
  \noindent #3\smallskip\\%
}

\begin{document}

\title{Mixed-precision GPU algorithms for efficient turbulent flow simulations with Raviart--Thomas finite elements}

\author{Ivan Prusak\thanks{Ruhr University Bochum, Universit\"atsstr. 150, 44801 Bochum, Germany (\texttt{\{ivan.prusak,enes.soydan,martin.kronbichler\}@rub.de})} \and
Enes Mustafa Soydan$^1$ \and
Ivan Pribec\thanks{Leibniz Supercomputing Centre, Boltzmannstr. 2, 85748 Garching b. M\"unchen, Germany, (\texttt{ivan.pribec@lrz.de})} \and
Martin Kronbichler$^1$
}
            
\allowdisplaybreaks
\maketitle

\begin{abstract}
We propose GPU algorithms for high-fidelity simulation of incompressible turbulent flows. Discretization in space is performed with $H(\text{div})$-conforming high-order Raviart--Thomas finite elements for the velocity and an $L^2$-conforming discontinuous Galerkin approximation for the pressure. In time, a consistent splitting scheme based on higher-order BDF time stepping is used, with convection treated explicitly. In this scheme, a pressure Poisson equation and a symmetric reaction-diffusion-type equation for the velocity need to be solved in each time step. We develop a solution framework with fast matrix-free operator evaluation for all ingredients, combined with multigrid solvers for the Poisson problem, and propose a robust mixed-precision algorithmic framework. A key to mixed-precision efficiency is a least-squares projection to generate accurate initial guesses for the iterative linear solvers, enabling us to work with relative residual tolerances of  $10^{-3}$. In this regime, running the solvers entirely in single precision leads to almost no change in overall iteration counts and maintains the crucial turbulence statistics, while showing up to $1.7\times$ speedup over pure double-precision simulations.
\end{abstract}

\noindent \textbf{Key words.} {
Raviart--Thomas elements,  matrix-free algorithm, sum factorization, GPU programming, mixed-precision arithmetic}

\section{Introduction}
\label{sec:introduction}

The accurate simulation of fluid flows demands a high resolution in both space and time,
especially in the case of turbulence,
where non-stationary small-scale structures develop. The range of relevant turbulent structures increases with the Reynolds
number~\cite{Pope00turbulent}. Many technical flows involve Reynolds numbers
where the full resolution of all physically relevant scales, a so-called
direct numerical simulation, would be well beyond the capabilities of even the
largest supercomputers in use today. Several modeling approaches exist to compute flows with lower effort, where
large-eddy simulation and Reynolds-averaged Navier--Stokes equations are
the most common ones. This work aims to increase the computational efficiency for under-resolved
flows in the context of implicit large eddy simulation, i.e., simulations where the
spatial and temporal resolution of the numerical scheme can only resolve
the larger of the relevant scales, while the physical dissipation happening at subgrid
scale is represented by numerical dissipation near the resolution limit~\cite{Fehn19high,Moura17eddy}.

A mathematical model for fluid flows is given by the Navier--Stokes equations.
For flows at low Mach numbers, i.e., with flow speeds well below the fluid's
speed of sound, compressibility is negligible and
allows to avoid resolving the fast acoustic time scales. After
discretization of the incompressible Navier--Stokes equations in space, a system of nonlinear ordinary differential equations
is obtained, which is discretized in time with a suitable time
integrator. At the end of this process, one or several linear systems of
equations need to be solved in each time step. With efficient implementations, especially on contemporary parallel and accelerator-based hardware, the main
computational effort is spent in the solution of these linear systems. The
main focus of this study is thus on linear solvers. We utilize robust
high-order non-conforming finite element techniques, which are crucial in the
aforementioned under-resolved context. Non-conforming discontinuous Galerkin
schemes have been demonstrated to provide excellent resolution
capabilities~\cite{Moura15linear}, to be applicable to complex geometries with
unstructured meshes~\cite{Fischer22nekrs,Hesthaven08nodal,Orszag80spectral},
and to match well with the arithmetic capabilities of modern hardware, where
access to global memory is expensive compared to arithmetic work on cached
data.

Against the background of many algorithmic and implementation contributions
during the past decades, the high-performance computing community has
shifted the focus towards using lower-precision
arithmetic compared to the traditional double precision (FP64) standard. There
are two main motivations to use a reduced precision, namely (a) the reduction
in memory transfer of narrower data types and (b) a possibly higher arithmetic
capability of hardware for those narrow types. For example, with its tensor
cores, a contemporary NVIDIA GH200 GPU provides $16\times$ the arithmetic
throughput for the half-precision FP16 type compared to FP64. There is a vast
amount of literature on mixed-precision algorithms in scientific computing and
numerical linear algebra, as illustrated, e.g., by two survey
works~\cite{Abdelfattah21survey,Higham22mixed,Kashi26mixed}. In fluid dynamics, the impact
of reduced precision was studied in several works, where results for
compressible flows with explicit schemes~\cite{Siklosi26,Witherden20impact}
with single-precision (FP32) format were encouraging, while results for
incompressible flows showed potential, but also
challenges for certain turbulent quantities, see~\cite{Delorme24,Karp26effects} and references
therein. In incompressible flows and with implicit time
integrators, the solution of linear systems is a crucial component, allowing the use of generic
linear algebra concepts~\cite{Luszczek2024batched}. Another research direction considering
higher-order discontinuous Galerkin discretizations specifically is the compression
of the solution vectors with advanced
block-floating point formats~\cite{Sundriyal26customized}, but the application to sophisticated turbulence solvers is yet to be done.

This work contributes to mixed-precision algorithms by proposing specialized
algorithms for cutting-edge high-order solvers for partial differential
equations as follows:
\begin{itemize}
\item We propose GPU algorithms for high-order Raviart--Thomas finite elements
  for spatial discretization, which have excellent accuracy and stability
  properties. However, they are non-conforming for the Navier--Stokes
  equations and thus require the use of face integrals; together with Piola
  transformations the implementation is more
  elaborate~\cite{Boffi13mixed,Pazner25subspace,wik22}, and
  hardware-aligned matrix-free solvers for Raviart--Thomas elements have so far only been 
  developed in the conforming case~\cite{Pazner23gpu,Pazner24matrixfree}. For time
  integration, we use a consistent splitting
  scheme~\cite{Guermond06overview,Liu09open,Still26discontinuous} to split the
  saddle-point velocity--pressure system into two positive definite
  sub-systems. While this leads to a small violation of the point-wise
  divergence-free condition up to third-order-in-time, the overall robustness
  of the method is retained. The main motivation is the high efficiency of the
  constituents of splitting schemes~\cite{Creff25}, despite recent progress on
  saddle point solvers~\cite{Benzi26}.
\item To overcome the memory-bandwidth limitation of classical finite-element
  workflows that first assemble a sparse matrix that is later handed to a
  linear algebra package for solution, we propose a matrix-free strategy that
  evaluates the finite-element integrals on the fly with fast quadrature,
  using sum-factorization techniques on hexahedral
  elements~\cite{Deville02high,Kronbichler12generic,Melenk01fully,Orszag80spectral}.
  We develop an implementation of these steps for GPUs and evaluate the main
  characteristics for iterative solvers on recent NVIDIA and AMD devices.
\item For increased efficiency, we propose a mixed-precision solution strategy
  that combines the concept of iterative refinement with high-quality initial
  guess computations for the linear solvers. This enables the use of FP32
  arithmetic for the main solver steps, while achieving tolerance levels below
  $10^{-9}$ as in FP64 arithmetic. The advantages of lower precision on
  matrix-free evaluation and multigrid solvers are demonstrated on GPUs.
\end{itemize}
The implementations are based on the infrastructure of the widely used deal.II finite element library, including support for distributed meshes, matrix-free algorithms for elements of arbitrary order, and capabilities for high-order curved boundaries~\cite{dealii97,dealiigeneric}. Although we consider an application in incompressible flows and specialize the algorithms in this setting, we emphasize that many of
the concepts are generic and also applicable to related time-dependent
partial differential equations. 

The remainder of this work is structured as follows. In section
\ref{sec:discretization} the discretization of the incompressible
Navier--Stokes equations with Raviart--Thomas elements is derived. Section~\ref{sec:solver_algorithms} details the mixed-precision approach to solving the associated linear systems. Section~\ref{sec:gpu} describes the main details of the GPU
implementation of matrix-free operator
evaluation, analyzes its performance on GPU systems, and presents the main
ingredients of multigrid solvers for the pressure Poisson equation. Section~\ref{sec:mixed_precision} evaluates the properties of the mixed-precision solver and the achieved speedup for an exemplary test case of turbulent incompressible flows. Section
\ref{sec:summary} contains a summary of the work performed and of the experimental results.

\section{Discretization of the incompressible Navier--Stokes equations}\label{sec:discretization}

Incompressible, viscous fluid flows are modeled by the Navier--Stokes equations, a system of partial differential
equations in the velocity $\mathbf{u}$ and pressure $p$ solved on a bounded
domain $\Omega\subset \mathbb{R}^{d}$, $d=2,3$,
\begin{equation}\label{eq:ns}
\begin{aligned}
  \frac{\partial \mathbf{u}}{\partial t} + \mathbf{u} \cdot \nabla \mathbf{u}
  - \nu \nabla^2 \mathbf{u} + \nabla p &= \mathbf{f},
  \\
  \nabla \cdot \mathbf{u} = 0,
\end{aligned}
\end{equation}
where $\nu$ denotes the fluid viscosity and $\mathbf{f}$ external forces. The
equation is complemented by boundary conditions (such as no slip, inflow, or outflow) and an initial velocity $\mathbf u(\cdot, 0)=\mathbf{u}_0$. The symbol
$\nabla$ denotes the gradient of a scalar or vector field with partial derivatives along the coordinate directions, $\nabla \cdot$ the
divergence of a vector field, and $\nabla^2 = \nabla \cdot \nabla$ the
Laplacian.

\subsection{Time integration}

A common approach to advance the Navier--Stokes equations in time is to use
projection-type methods, i.e., to substitute some of the variables in the coupled
velocity--pressure system in order to obtain separate systems for velocity
and pressure~\cite{Guermond06overview}. In this work, we choose the consistent
splitting method proposed in~\cite{Liu09open}, which is based on the BDF time
step family of order $J$, and follow the recent algorithmic realization
by~\cite[Sec.~3.3]{Still26discontinuous}, with the exception of using an explicit
formulation for the convective term. When stepping to time level $n+1$, the
method performs two main steps,
\begin{itemize}
\item a pressure step for the (modified) pressure,
\begin{equation}\label{eq:ns_pressure}
  -\nabla^2 p^{n+1} = \sum_{i=1}^{J_c} \hat{\beta}_i \nabla\cdot
  \left(\mathbf{u}^{n+1-i}\cdot \nabla \mathbf{u}^{n+1-i}\right)
  - \nabla \cdot \mathbf{f}^{n+1}
  - \sum_{i=1}^J \frac{\alpha_i}{\Delta t} \nabla \cdot \mathbf{u}^{n+1-i},
\end{equation}
with right-hand side given by extrapolations of convection, the body force and
previous-time Leray projection terms \cite{Liu09open} combined with suitable
boundary conditions not stated here for brevity~\cite[Sec.~3.3]{Still26discontinuous}, and
\item a momentum step subject to the update formula
\begin{equation}\label{eq:ns_momentum}
  \frac{\alpha_0}{\Delta t} \mathbf{u}^{n+1} - \nu \nabla^2 \mathbf{u}^{n+1} =
  \sum_{i=1}^J \left(\frac{\alpha_i}{\Delta t} \mathbf{u}^{n+1-i}\right)
  -\sum_{i=1}^J \left(\beta_i \mathbf{u}^{n+1-i} \cdot \nabla \mathbf{u}^{n+1-i}\right)
   - \nabla p^{n+1}
  + \mathbf{f}^{n+1} .
\end{equation}
\end{itemize}
In these equations, $\Delta t$ denotes the time step size and $(\cdot)^{n+1}$
the time level, the factors $\alpha_i$, $i=0,1,\ldots,J$ the coefficients in the BDF time
integrator of order $J$, while $\beta_i, i=1,\ldots,J $ and $\hat{\beta}_i, i=1,\ldots J_c$ are the coefficients of the associated extrapolation methods of orders $J$ and $J_c$, respectively. The
extrapolation order of the pressure right-hand side $J_c = J-1$ is
chosen for its improved stability~\cite{Still26discontinuous}.

\subsection{Spatial discretization with finite elements}


For representing spatial variables, we use an unstructured mesh of possibly deformed
hexahedral cells $\Omega =\bigcup_{e=1}^N \Omega_e$ of characteristic size $h$.
On each cell, polynomials approximate the solution for pressure, denoted by
$p_h$, and velocity, denoted by $\mathbf u_h$, with an expansion in terms of
basis functions and unknown
coefficients. For velocity, we choose $k$-th order tensor-product
Raviart--Thomas finite elements, which use an
anisotropic polynomial space of degree $k+1$ in the normal direction for
the respective velocity component and degree $k$ in the tangential
direction(s). Continuity is imposed in normal direction whereas the function
space is discontinuous tangentially, making the space $H(\text{div})$-conforming~\cite{Boffi13mixed}. Combined with a
discontinuous pressure space of degree $k$, the resulting solution satisfies the condition
$\nabla \cdot \mathbf{u} = 0$ pointwise for the Navier--Stokes
system~\eqref{eq:ns}, rather than only up to discretization errors in
conventional schemes. This makes the method very robust for
challenging flows, in particular in the regime of under-resolved flows~\cite{Fehn19high}.

Since the velocity function space is non-conforming with respect to the spatial derivatives in~\eqref{eq:ns}, and the pressure approximation is non-conforming for the approximation of the Poisson equation~\eqref{eq:ns_pressure},
a discontinuous Galerkin formulation including face integrals is
necessary to devise a scheme consistent with the underlying differential equation~\cite{Hesthaven08nodal}. We use a
symmetric interior penalty formulation~\cite{Arnold02unified} for the viscous
and pressure Poisson terms, an upwind-like discretization for convection and
the $H(\text{div})$-conforming expressions for the pressure gradient and velocity
divergence.

We denote by $(\cdot, \cdot)_{\Omega_e}$ the bilinear form of integrals of the dot product of two scalar, vector or tensor
quantities on the
domain $\Omega_e$ in the usual sense, and by
$\left<\cdot, \cdot\right>_{\partial \Omega_e}$ the integral over the
boundary of $\Omega_e$. We denote by $\jump{\mathbf v} = \mathbf v^- - \mathbf v^+$ the jump
of a quantity $\mathbf v$ across two sides ${}^-$ and ${}^+$ of a face and by
$\average{\mathbf v} = \frac 12 \left(\mathbf v^- + \mathbf v^+\right)$ the
average. At the boundary of $\Omega$, suitable definitions of the exterior value
$\mathbf v^{+}$ are used to impose boundary conditions via the mirror
principle~\cite{Hesthaven08nodal}. The unit outer normal vector on the
boundary of $\Omega_e$ is denoted by $\mathbf n$. The final weak form
of~\eqref{eq:ns_pressure} finds the unknown coefficients in the discrete pressure $p=p^{n+1}_h$ such that for all
elements $\Omega_e$ and all test functions $q\in \mathbb{Q}_k$, there holds
\begin{equation}\label{eq:pressure_weak}
  \left(\nabla q, \nabla p
  \right)_{\Omega_e} 
  - 
  \left<\nabla q \cdot\mathbf{n}, \frac 12 \jump{p}\right>_{\partial \Omega_e}
  - 
  \left<q, \average{\nabla p}\cdot \mathbf n - \tau_p \jump{p}\right>_{\partial \Omega_e}
  = r_{p,\Omega_e} (q).
\end{equation}
Here, $\tau_p$ is a penalty factor chosen approximately as $\frac {k^2}{h}$
for degree $k$ and element size $h$ to ensure coercivity of the
discretization through an inverse estimate, see, e.g., \cite{Hesthaven08nodal}
and references therein for a precise definition. The right-hand side term
$r_{p,\Omega_e} (q)$ collects the weak forms of the discretization of
right-hand side terms in Eq.~\eqref{eq:ns_pressure}
following~\cite{Still26discontinuous}. Note that the discretization of the convective right-hand side also includes face
integrals to account for the missing $H^1$-conformity, while the divergence
terms $\nabla \cdot \mathbf u^{n+1-i}$ are discretized in a conforming way,
i.e., avoiding face integrals. Likewise, the weak form of Eq.~\eqref{eq:ns_momentum}
finds the polynomial coefficients describing the velocity field $\mathbf u = \mathbf u^{n+1}_h$ such that for all
elements $\Omega_e$ and all test functions $\mathbf v$ there holds
\begin{equation}\label{eq:velocity_weak}
  \frac{\alpha_0}{\Delta t} \left(\mathbf v, \mathbf u\right)_{\Omega_e} + \nu \left[ \left(
      \nabla \mathbf {v}, \nabla \mathbf{u}
    \right)_{\Omega_e}
    -
    \left<\nabla \mathbf{v}\cdot\mathbf{n}, \frac 12 \jump{\mathbf{u}}\right>_{\partial \Omega_e}
    -
    \left<\mathbf{v}, \average{\nabla \mathbf{u}}\cdot \mathbf n -
      \tau \jump{\mathbf{u}}\right>_{\partial \Omega_e}\right]
  = r_{\mathbf{u},\Omega_e} (\mathbf{v}),
\end{equation}
again denoting by $r_{\mathbf{u},\Omega_e} (\mathbf{v})$ the right-hand side
terms of Eq.~\eqref{eq:ns_momentum} evaluated according to the usual
discontinuous Galerkin procedure~\cite{Still26discontinuous}. For the nonlinear
convective term, a Gaussian quadrature formula with
$\left \lfloor \frac{3 k}{2} \right\rfloor + 1$ points per direction is used to
ensure consistent integration (over-integration). An important
application requirement is to ensure good performance over a range of
polynomial degrees, $1 \leq k \leq 8$, enabling the optimal selection of mesh
resolution and polynomial degree depending on the geometry and
flow features.

\section{Mixed-precision strategy for solving linear systems}\label{sec:solver_algorithms}

Assuming that the polynomial expansion coefficients on all elements for the discrete solution fields $p^{n+1}_h$ and $\mathbf{u}^{n+1}$
are denoted by $P^{n+1}$ for pressure and
$U^{n+1}$ for velocity, solving for the weak
forms~\eqref{eq:pressure_weak}--\eqref{eq:velocity_weak} on all elements leads
to two linear systems,
\begin{align}
  \mathbf{L} P^{n+1} &= B^{n+1}, \label{eq:pressure_lin}\\
  \left(\frac{\alpha_0}{\Delta t} \mathbf{M} + \nu \mathbf{K} \right) U^{n+1} &= R^{n+1}, \label{eq:velocity_lin}
\end{align}
where $\mathbf{L}$ denotes a Poisson matrix for the pressure, $\mathbf{M}$ the velocity
mass matrix, $\mathbf{K}$ the viscous matrix, a vector-valued Poisson matrix, and
$B^{n+1}$ and $R^{n+1}$ are the respective right-hand side terms of the linear
systems. Both systems are symmetric and positive definite, with the non-linear,
non-symmetric convective effects treated explicitly in time.

\subsection{Solvers and preconditioners}

The pressure Poisson equation~\eqref{eq:pressure_lin} involves a matrix
  with global domain of influence and eigenvalue spectrum distributed as
  $k^4 h^{-2}$. To solve this system, multigrid methods have been established
  as one of the most effective
  strategies~\cite{Gholami16fft,Kanschat04multilevel,Trottenberg01multigrid}. In combination with
  matrix-free operator evaluation and suitable smoothers, they have been shown
  to deliver excellent performance with optimal $\mathcal O(n)$ complexity for
  systems of size $n$, including scalability to large
  supercomputers~\cite{Arndt20exadg, Bauer2020TerraNeo,Kronbichler18performance,Munch23efficient,Rudi2015EarthMantle}. To
  maximize efficiency, we use a multigrid V-cycle as a preconditioner within a conjugate gradient
  solver~\cite{Sundar15,Trottenberg01multigrid}. Further details on the multigrid scheme are given in Sec.~\ref{sec:multigrid} below.

The momentum equation~\eqref{eq:velocity_lin} is dominated by the mass
  matrix in the case of turbulence, since the Courant--Friedrichs--Lewy (CFL) condition on the convective
  terms necessitates $\Delta t \lesssim \frac{h}{k^{1.5} \|\mathbf{u}\|}$ and
  solvers are run in barely resolved cases, i.e.,
  $h \gtrsim \frac{\nu}{\|\mathbf{u}\|} $. As a consequence, the overall condition
  number is $\mathcal O(1)$, with a conjugate gradient solver using a preconditioner tailored for the mass matrix being a frequent choice~\cite{Krank17high}. We
  found that a point-Jacobi preconditioner, i.e., the inverse of the diagonal of the matrix
  $\frac{\alpha_0}{\Delta t} \mathbf{M} + \nu \mathbf{K}$, gives the best
  balance between the competing goals of preconditioner cost and iteration
  counts: With 2--3 iterations, a residual reduction of a factor 10 can be
  reached for typical mesh resolutions and time step sizes.

We choose the time step size adaptively to fulfill a CFL limit on each cell $\Omega_e$.
The variation in $\Delta t$ from one step to the next means that
the diagonal would need to be recomputed in every step if done separately,
which we address by computing $\text{diag}(\mathbf M)$ and $\nu\, \text{diag}(\mathbf{K})$
as two separate vectors during setup and combining them in each time step with the respective factors.

\subsection{Projective computation of an initial guess}\label{sec:initial_guess}

For small time steps, the solutions $P^{n}$ and $U^{n}$ are similar in
adjacent time steps. Thus, linear combinations of
$P^{n}, P^{n-1}, \ldots, P^{n+1-m}$ for some integer $m$ can provide a
good initial guess for the linear system in $P^{n+1}$, and similarly
for $U^{n+1}$.  Our approach builds on the idea by
Fischer~\cite{Fischer98projection}, see, e.g.,
\cite{Austin21initial,Wells26wing} for an overview of the main approaches and
some recent advances. Following the projection approach
by~\cite{Lohner04projective}, we find the initial guess
$P^{n+1}_0 = \sum_{i=1}^m \gamma_i P^{n+1-i}$ as the vector minimizing the
residual of $\mathbf{L} P^{n+1}_0 = B^{n+1}$ for the coefficients $\gamma_i$,
$i=1,\ldots,m$. We hence solve the least-squares problem
\begin{equation}\label{eq:projection}
  \min_{\gamma_i} \left \|
    \left(\mathbf{L} \sum_{i=1}^m\gamma_i P^{n+1-i}\right) - B^{n+1}\right\|_2
  =
  \min_{\gamma_i} \left \|
    \left(\sum_{i=1}^m \left(\mathbf{L}  P^{n+1-i}\right) \gamma_i\right) - B^{n+1}\right\|_2.
\end{equation}
Since $\mathbf{L} P^{n+1-i}$ is the left-hand side of previous linear systems, it is substituted
by the vectors $B^{n+1-i}$. The size $m$ is determined by balancing two competing goals, as
more vectors improve the quality and the residual reduction, but each added
vector increases the memory transfer in a memory-bound operation to solve
Eq.~\eqref{eq:projection}. Based on
extensive tests including modified Gram--Schmidt and QR orthogonalization
methods, we found similarly to~\cite{Wells26wing} that the best performance
for moderate $m\leq 6$ is the solution of
the associated normal equations in a windowed fashion with $m$ vectors,
\[
  Y^\top Y \gamma = Y^\top B^{n+1},
\]
where the $n\times m$ matrix $Y$ contains the columns $B^{n+1-i}$,
$i=1,\ldots, m$. The small $m\times m$ matrix $Y^\top Y$ is solved with a Cholesky factorization.

For the momentum equation, the overall approach is similar, with one key
modification. Due to the variable time step size, which is used to maximize the
efficiency with respect to the CFL stability limit, the
effective matrix $\frac{\alpha_0}{\Delta t} \mathbf{M} + \nu \mathbf{K}$
differs between time steps. Therefore, the columns in $Y$ cannot be recycled
from right-hand sides of previous time steps as in the case of $\mathbf{L}$. Instead of
re-computing $m$ columns in each time step or accept a lower accuracy when assuming
that the matrix does not change~\cite{Lohner04projective}, we opt to store the
old right-hand sides $Y_\text{R}(:,i) = R^{n+1-i}$ and the matrix-vector
product with mass matrix $Y_\text{M}(:,i) = \mathbf{M} U^{n+1-i}$ separately
and combine the two vectors while evaluating $Y^\top Y$ and $Y^\top R^{n+1}$ to
compensate for the time-step difference, without writing back updated vectors.
As these operations are limited by
memory bandwidth and each vector $i$ is accessed once per time step, this
approach minimizes the overall memory access.

\begin{algorithm}
  \caption{Mixed-precision least-squares algorithm for computing an initial guess.}
  \label{alg:projective}
  \begin{tabularx}{\linewidth}{>{\raggedright\arraybackslash}p{1.3em} X}
    \AlgStep{lowPrec}{1}{%
      Rotate columns of $Y$, $Y(:,i+1) \leftarrow Y(:,i)$ for
      $i=1, \ldots, m$. \colorbox{white}{No memory access.}
    }
    \AlgStep{lowPrec}{2}{%
      Compute the dot products of the symmetric matrix $C = Y^\top Y$  stored in (LP) and the
      right-hand side by a fused kernel, doing one single
      memory access for each
      vector. For the velocity equation, subtract suitable factors of $Y_M$ from $Y_R$
      to form $Y$ on the fly. \colorbox{highPrec}{Accumulations are done in (HP).} \colorbox{white}{Memory access:
      $m n \cdot \texttt{sizeof}(\text{LP}) + n \cdot \texttt{sizeof}(\text{HP})$.}
    }
    \AlgStep{highPrec}{3}{%
      Solve $m\times m$ system of normal equations for $\gamma$ by Cholesky
      factorization in (HP). If the diagonal entry  $C_{\ell,\ell}$ is below a threshold
      $\epsilon^2 = 10^{-20}$, set $\tilde{m} = \ell-1$, otherwise use
      $\tilde{m} = m$.
    }
    \AlgStep{lowPrec}{4}{%
      Compute $P^{n+1}_0 = (P^{n}, P^{n-1},\ldots, P^{n+1-\tilde{m}})\gamma$
      in (LP) and copy $B^{n+1}$ to (LP) as $Y(:,m+1) = B^{n+1}$ for next time step.
    }
  \end{tabularx}
\end{algorithm}

Since the intent with the initial guess computation is to reduce the initial
residual $B^{n+1} - \mathbf{L} P^{n+1}_0$ an iterative solver has to start
from, but does not otherwise influence the application accuracy, it is a
candidate to be run in a lower precision, motivated by compressed-basis GMRES
results~\cite{Aliaga22compressed}. Algorithm~\ref{alg:projective} describes the
steps in the algorithm execution and the associated precisions in a
two-precision simulation, using numbers of low precision (LP) and high
precision (HP). In this work, we use FP32 as LP and FP64 as HP, but other
combinations are also conceivable, given adequate exponent range and mantissa
digits to represent the unavoidable cancellation effects in a least-squares
system.

\subsection{Mixed-precision solution algorithm}

With the ingredients described in the two previous subsections, we can now
formulate an algorithm that can run the largest share of operations within a
time step in lower precision. Our approach is related to the concept of
iterative refinement~\cite{Higham22mixed}, where we require solvers to
only provide a modest reduction in the residual.
Our contribution is aligned with the properties of an optimized computational fluid dynamics (CFD) solver:
\begin{itemize}
\item the use of matrix-free operator evaluation makes the cost of a
  matrix-vector product similar to that of a finite difference stencil or a vector update as soon as the data exceeds fast cache memory, meaning that vector operations are significant for overall cost, see, e.g., \cite{Kronbichler23enhancing};
\item preconditioning is very cheap for the momentum
  equation~\eqref{eq:velocity_lin}, while the multigrid V-cycle in the
  pressure Poisson solver is a rather heavy preconditioner; the algorithm components done on
  each multigrid level resemble the steps in the momentum equation;
\item due to optimal-complexity solvers, the iteration counts are relatively
  low in the regime of 2 (Poisson) to 10 (momentum), as shown in the results section
  below. The weak forms involved in the right-hand side terms can be evaluated
  at similar performance as a matrix-vector product.
\end{itemize}

A crucial ingredient in our solver is the initial guess computation from
Algorithm~\ref{alg:projective}. It provides a high-quality starting vector for
the iterative solvers, reducing the residual by around $10^4$ for
the pressure Poisson equation~\eqref{eq:pressure_lin} and by $10^6$ for the
momentum equation~\eqref{eq:velocity_lin}, compared to the norm of the
right-hand side. Hence, any iterative solver whose number format precision suffices to reduce
the unpreconditioned residual norms by $10^2$ to $10^3$ beyond the starting
residual is applicable. Concretely, we propose to run the entire solver for both
Eq.~\eqref{eq:pressure_lin} and Eq.~\eqref{eq:velocity_lin} in lower
precision, avoiding an inner-outer nesting of solvers, which allows Krylov
subspace solvers to maintain a single search space. Algorithm~\ref{alg:mixed_precision}
presents the details of our mixed-precision CFD algorithm.

\begin{algorithm}
  \caption{Mixed-precision algorithm to advance one time step.}
  \label{alg:mixed_precision}
  \begin{tabularx}{\linewidth}{>{\raggedright\arraybackslash}p{1.3em} X}
    \AlgStep{lowPrec}{1}{%
      Compute time step size $\Delta t$ from $U^{n}$ (LP).
    }
    \AlgStep{highPrec}{2}{%
      Evaluate convective terms for both the weak forms
  arising from
  $\hat{\beta}_i \nabla\cdot \left(\mathbf{u}^{n}\cdot \nabla
    \mathbf{u}^{n}\right)$ for the pressure right-hand
  side~\eqref{eq:ns_pressure} as well as
  $\beta_i \mathbf{u}^{n}\cdot \nabla \mathbf{u}^{n}$ for the momentum right-hand
  side~\eqref{eq:ns_momentum} in a single integration loop with different test
  functions using (HP).
}
\AlgStep{highPrec}{3}{%
  Compute right-hand side of pressure, including combination of
  convective contributions from $J_c$ previous time steps (HP).
}
\AlgStep{lowPrec}{4}{%
  Compute pressure initial guess $P^{n+1}_0$ with Alg.~\ref{alg:projective} in (LP).
  }
  \AlgStep{lowPrec}{5}{%
    Solve pressure Poisson system~\eqref{eq:pressure_lin} with conjugate
    gradient algorithm in (LP) preconditioned by multigrid (LP). \colorbox{highPrec}{Use
    (HP) as fallback if convergence stagnates.}
  }
  \AlgStep{highPrec}{6}{%
    Compute momentum right-hand side, including summations of convective
    terms from $J$ previous time steps (HP).
  }
  \AlgStep{lowPrec}{7}{%
    Compute initial guess $U_0^{n+1}$ with Alg.~\ref{alg:projective} in (LP).
  }
  \AlgStep{highPrec}{8}{%
    Compute linear residual
\[
  \widetilde{R}^{n+1} = R^{n+1} - \left(\frac{\alpha_0}{\Delta t} \mathbf{M} +
    \nu \mathbf{K} \right) U^{n+1}_0 \qquad \qquad \text{(HP).}
\]
  \vspace{-0.3cm}
  }
  \AlgStep{lowPrec}{9}{%
    Prepare diagonal point-Jacobi preconditioner from the two precomputed
  vectors
  $D^{n+1} = 1 / \left( \frac{\alpha_0}{\Delta t} \text{diag}(\mathbf{M}) +
    \text{diag}(\nu \mathbf{K})\right)$ (LP).
  }
  \AlgStep{lowPrec}{10}{%
    Run preconditioned conjugate gradient solver for Eq.~\eqref{eq:velocity_lin} with r.h.s. $\widetilde{R}^{n+1}$ in (LP).
  }
  \AlgStep{highPrec}{11}{%
    Update $U^{n+1}$ in (HP), \colorbox{lowPrec}{store LP copy}.
    }
  
  \AlgStep{highPrec}{12}{%
    Post-processing steps, like turbulence statistics or visualization
    preparation (HP).
  }
\end{tabularx}
\end{algorithm}

Our approach is to select solver tolerances such that the main turbulence
results are unaffected, taking a more systematic approach than the
works~\cite{Karp26effects,Witherden20impact} which merely reported the impact of
lower precision on results. For many turbulent cases that require small time
steps, such as the one presented in Sec.~\ref{sec:mixed_precision}, we found
that residual reductions of at least $10^8$ compared to the right-hand side
norm were necessary for the velocity, and $10^4$ or $10^5$ for the
pressure. Given the limited precision of the LP path, we therefore track the
residual reduction progress of our solver with the following strategy:
\begin{itemize}
\item For the pressure Poisson equation in step 5 of
  Algorithm~\ref{alg:mixed_precision}, we first aim for a residual reduction
  of at most $10^4$ (or smaller if the initial guess is good enough), which is
  a regime where the chosen LP format FP32 gives confident results. If the
  result is insufficient, a (HP) residual is computed and an outer CG solver
  run in HP (i.e., FP64) is used. Combining lower FP32 precision in a
  multigrid V-cycle with a higher-precision outer solver in FP64 is capable of
  a relative residual reduction down to at least $10^{-12}$, as originally
  suggested by~\cite{Gropp00performance}, later quantified in the matrix-free
  case by \cite{Bauer2020TerraNeo,Kronbichler12generic,Kronbichler19gpu} and recently embedded
  in a much wider theory and ingredient setting by~\cite{Vacek2026Mixed}, see also
  references therein.
\item For the momentum equation, we interrupt the LP solver once the reported CG residual
  reduction exceeds $10^5$ or iteration counts exceed 20. We then repeat steps
  8 and 10 of Algorithm~\ref{alg:mixed_precision}, computing a HP residual and
  solving in LP again.
\end{itemize}
In our computations we never observed this fall-back to be necessary with the
exception of the first two time steps, when the projective guess does not yet
have enough search directions and the residual reduction is higher.

\subsection{Target Architectures and Software Implementation}

We use an implementation of the algorithms with the deal.II
library~\cite{dealii97,dealiigeneric}, in particular the matrix-free
infrastructure presented
in~\cite{Kronbichler12generic,Kronbichler19fast,Kronbichler19gpu} and
multigrid methods~\cite{Clevenger21flexible,Kanschat04multilevel,Kronbichler18performance,Kronbichler19gpu,Munch23efficient}. 
As demonstrated in~\cite{Kronbichler18performance}, matrix-free algorithms can
deliver around an order of magnitude faster time to solution than matrix-based implementations for moderate polynomial
degrees $k=3,4$ due to a much reduced memory access. Their
throughput, measured as degrees of freedom processed per second (DOF/s) during operator
evaluation or one solver iteration~\cite{Fischer20scalability}, is also several times higher than for continuous finite elements of degree $k=1$ with sparse matrices.
For the GPU algorithms, we use the Kokkos framework~\cite{kokkos} and its
interfaces in deal.II, but with custom-built kernels to reach the
performance capabilities of the GPU
hardware reported in~\cite{Abdelfattah21gpu,Kolev21efficient}. Experiments were performed on two GPU platforms: Operator evaluation tests are conducted on a single die of an AMD MI250X GPU (1.3 TB/s stream bandwidth) at the LUMI supercomputer and on an NVIDIA GH200 GPU (3.6 TB/s stream bandwidth) hosted on the JUPITER supercomputer. 
The subsequent multigrid tests were performed exclusively on the NVIDIA GH200 platform. For distributed-memory execution, communication is handled via MPI, mapping one GPU per MPI rank.

\section{GPU implementation of matrix-free operator evaluation}\label{sec:gpu}

The core computational kernel for matrix-free operator evaluation is to compute the finite-element integrals on the fly. For tensor product shape functions evaluated with a tensor-product quadrature formula, as typical on three-dimensional hexahedral meshes, the main step is to compute evaluations of polynomial interpolations or derivatives at quadrature points, or summations over quadrature points for test functions, which can be expressed by a sequence of tensor contractions~\cite{Abdelfattah21batched,Kolev21efficient,Kronbichler12generic, Kronbichler19fast},
$$
u^e_{abc} = \sum\limits_{k}  I_{ck} \sum\limits_{j} I_{bj} \sum\limits_{i} I_{ai} u^e_{ijk},
$$
where $u^e_{ijk}$ represents the scalar field values at the finite element nodes for element $e$ and $I$ denotes the 1D interpolation matrix. The indices $i, j, k$ specify the nodal coordinates, while $a, b, c$ refer to the corresponding quadrature point indices along each spatial dimension.
An efficient distribution of this workload across the GPU grid, block, and thread hierarchy was proposed by \cite{SwirydowiczHighorder}, where each element is assigned to a single thread block, see also~\cite{cui24acceleration,Kronbichler19gpu}. This approach  
delivers high throughput, particularly for high-order elements. However, for low-order elements, this strategy results in poor warp utilization and hardware occupancy. 
For instance, NVIDIA architectures impose a strict limit of 32 active thread blocks per Streaming Multiprocessor (SM) \cite{nvidiaHopperTuning}, which means that blocks with small thread counts cannot fully saturate the hardware. This bottleneck leaves the shared memory and register resources within each SM underutilized. Alternatively, as shown in \cite{EichstaedtGPU}, assigning one thread per element in 2D meshes 
leads to fast execution only for polynomial orders $k < 5$. To address these occupancy limitations and mitigate the performance sensitivity to the polynomial order, 
we introduce a batch-processing method for 3D hexahedral meshes in which multiple elements are mapped to a single thread block as detailed in Table~\ref{tab:gpu_mapping_compute}, akin to cross-element vectorization strategies proposed previously for CPUs~\cite{Kronbichler12generic,Kronbichler19fast}. This strategy mitigates underutilization in 
low-order configurations and delivers consistently high performance across all evaluated 
polynomial orders. The proposed batching framework can be interpreted as a generalization of the conventional single-element-per-block strategy; 
when the batch size is set to $N = 1$, the mapping naturally assigns a single element per thread block. 

\begin{table}
\centering
\caption{Mapping of tensor-contraction workloads onto the GPU execution hierarchy for vector units.}
\label{tab:gpu_mapping_compute}
\small
\begin{tabular}{lcl}
\toprule
\textbf{Execution Scope} & \textbf{Local Contraction} & \textbf{Workload Mapping} \\ 
\midrule
Grid  & $u^e_{ajk} = \sum\limits_{i} I_{ai} u^e_{ijk}$                     & 1D layout over element batch  \\ \nopagebreak
Block  & $u^{\tilde{e}}_{ajk} = \sum\limits_{i} I_{ai} u^{\tilde{e}}_{ijk}$ & 3D layout over $\tilde{e}, j, k$ \\ \nopagebreak
Thread & $u_{a} = \sum\limits_{i} I_{ai} u_{i} $                             & Matrix-vector product \\ 
\bottomrule
\end{tabular}
\end{table}

\begin{figure}
    \centering
    \begin{tikzpicture}
        
        \begin{axis}[
            name=ax1, 
            xlabel={Polynomial Order},
            ylabel={Elements per Thread Block},
            width=0.5\columnwidth,
            height=0.4\columnwidth,
            title style={font=\footnotesize},
            tick label style={font=\footnotesize},
            label style={font=\footnotesize},
            legend style={font=\footnotesize},
            grid=major,
            grid style={lightgray!30},
            xtick distance=1,          
            ymode=log,
            log basis y=2,
            log origin=infty,
            ymin=0.5, ymax=127,
            yticklabels={1,2,4,8,16,32,64,128},
            ytick={1,2,4,8,16,32,64,128},     
            ybar=2pt,                 
            bar width=6pt,
            xtick={1,...,8},            
            enlarge x limits=0.15,    
            nodes near coords={\pgfmathprintnumber{\pgfkeysvalueof{/data point/y}}},
            point meta=y,             
            every node near coord/.append style={
                font=\tiny,           
                yshift=0.5pt,           
                /pgf/number format/fixed,
                /pgf/number format/precision=0
            },
            legend pos=north east,
            legend style={draw=none, fill=white}
        ]

            \addplot[fill=blue!50, draw=blue!80] coordinates {
                (1,78) (2,23) (3,9) (4,5) (5,2) (6,1) (7,1) (8,1)
            };
            \addlegendentry{FP32}

            \addplot[fill=red!50, draw=red!80] coordinates {
                (1,39) (2,11) (3,4) (4,2) (5,1) (6,1) (7,1) (8,1)
            };
            \addlegendentry{FP64}
        \end{axis}

        \begin{axis}[
            name=ax2,
            at={(ax1.outer east)}, 
            width=0.5\columnwidth,
            height=0.4\columnwidth,
            title style={font=\footnotesize},
            tick label style={font=\footnotesize},
            label style={font=\footnotesize},
            legend style={font=\footnotesize},
            anchor=outer west,
            xshift=0.5cm, 
            xlabel={Polynomial Order},
            grid=major,
            grid style={lightgray!30},
            xtick distance=1,
            ymode=log,
            log basis y=2,
            log origin=infty,
            ymin=0.5, ymax=127,
            yticklabels={1,2,4,8,16,32,64,128},
            ytick={1,2,4,8,16,32,64,128},     
            ybar=2pt,                 
            bar width=6pt,
            xtick={1,...,8},            
            enlarge x limits=0.15,    
            nodes near coords={\pgfmathprintnumber{\pgfkeysvalueof{/data point/y}}},
            point meta=y,             
            every node near coord/.append style={
                font=\tiny,           
                yshift=0.5pt,           
                /pgf/number format/fixed,
                /pgf/number format/precision=0
            },
            legend pos=north east,
            legend style={draw=none, fill=white}
        ]

            \addplot[fill=teal!50, draw=teal!80] coordinates {
                (1,18) (2,7) (3,4) (4,2) (5,1) (6,1) (7,1) (8,1)
            };
            \addlegendentry{FP32}

            \addplot[fill=violet!50, draw=violet!80] coordinates {
                (1,9) (2,3) (3,2) (4,1) (5,1) (6,1) (7,1) (8,1)
            };
            \addlegendentry{FP64}
        \end{axis}
        
    \end{tikzpicture}
    \caption{Resulting elements per thread block to reach a 10\,kB shared memory allocation per block as a function of the polynomial degree. The comparison highlights the scaling differences between 
    (left) the scalar Laplace operator ($\mathbf{L}$) and (right) the vector Raviart-Thomas mass operator ($\mathbf{M}$).}
    \label{fig:batchCount}
\end{figure}
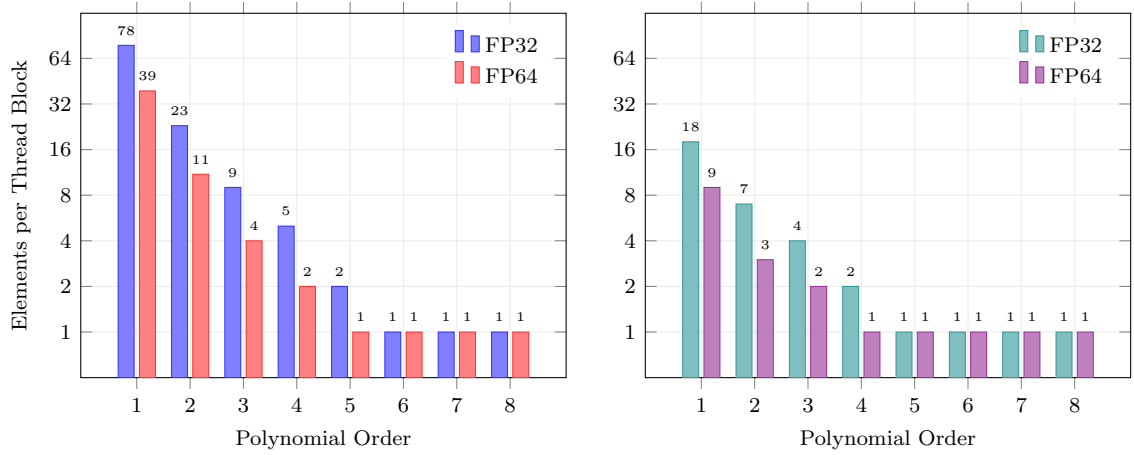

A critical consideration in this approach is determining the optimal element batch size required to achieve peak performance. To formalize this, the shared memory consumption per thread block can be modeled as a function $S_{\text{block}}(N, N_q, B)$, where $N$ is the element batch size, $N_q$ is the number of quadrature points per element, and $B$ represents the data precision in bytes. 
In practice, restricting $S_{\text{block}}$ to 10 kB allows the hardware scheduler to concurrently host multiple blocks without exceeding the maximum constraint of 32 thread blocks per Streaming Multiprocessor (SM).
As a concrete example, evaluating a 3D scalar Laplace operator like for Eq.~\eqref{eq:pressure_lin} yields the following equation for shared memory consumption per block:
$$
S_{\text{block}}(N, N_q, B) = N \cdot \left(4 N_q^3\right) \cdot B,
$$
where 4 is the number of arrays used per quadrature point (values, 3 components of gradient). For linear elements ($k=1$) with $N_q = 2$ quadrature points per dimension and FP32 precision ($B=4$), 
this equation can be solved for the maximum element batch size $N$ that fits within a targeted 10 kB ($10{,}000$ bytes) shared memory limit:
$$N = \max\left(\left\lfloor \frac{10{,}000}{4 \cdot 2^3 \cdot 4} \right\rfloor,1\right) = 78$$
Following this shared memory model, the full range of calculated batch sizes for the scalar Laplace operator across varying polynomial orders is presented in Figure~\ref{fig:batchCount}, alongside the corresponding calculations for the vector-valued Raviart--Thomas mass operator evaluation. Clearly, the lower precision arithmetic allows for more concurrency with more performance potential. Also, the vector-valued Raviart--Thomas operator with increased local sizes leads to smaller batches.

Beyond vector units, modern architectures offer specialized matrix units such as NVIDIA Tensor Cores \cite{nvidia_mma_instruction} or AMD Matrix Cores \cite{amd_matrix_calculator} 
that can be leveraged to exploit mixed-precision arithmetic for enhanced throughput and energy efficiency. In this work, we also present a novel technique to distribute our tensor contraction 
workload across specialized matrix units, as detailed in Table~\ref{tab:gpu_mapping_matrix}. The strategy relies on reshaping and transposing the tensor contraction $I_{ai} u^{\tilde{e}}_{ijk} $ into a flattened matrix format $u_{\tilde{e}kj,i} I_{i,a}$. By collapsing the batched element index ($\tilde{e}$) along with the spatial dimensions ($k, j$) into a single mode, 
the contraction over index $i$ is cast as a standard matrix-matrix multiplication (GEMM) that maps directly to the hardware matrix units.
Because matrix units operate on fixed-size tile dimensions, the resulting matrix-matrix multiplication is implemented using a hierarchical tiling strategy, where a single warp executes the entire tiled GEMM for the batch.

\begin{table}[htbp]
\centering
\caption{Mapping of tensor contraction workloads onto the GPU execution hierarchy for matrix units.}
\label{tab:gpu_mapping_matrix}
\small
\begin{tabular}{lcl}
\toprule
\textbf{Execution Scope} & \textbf{Local Contraction} & \textbf{Workload Mapping} \\ 
\midrule
Grid   & $u^e_{ajk} = \sum\limits_{i} I_{ai} u^e_{ijk}$                     & 1D layout over element batch  \\ \nopagebreak
Block  & $u^{\tilde{e}}_{ajk} = \sum\limits_{i} I_{ai} u^{\tilde{e}}_{ijk}$ & Single warp per block\\ \nopagebreak
Warp   & $u_{\tilde{e}kj,a} = \sum\limits_{i} u_{\tilde{e}kj,i} I_{i,a}$     & Tiled GEMM \\ 
\bottomrule
\end{tabular}
\end{table}

\subsection{Performance of operator evaluation}
\begin{figure}
    \centering
    \resizebox{0.9\textwidth}{!}{%
        \input{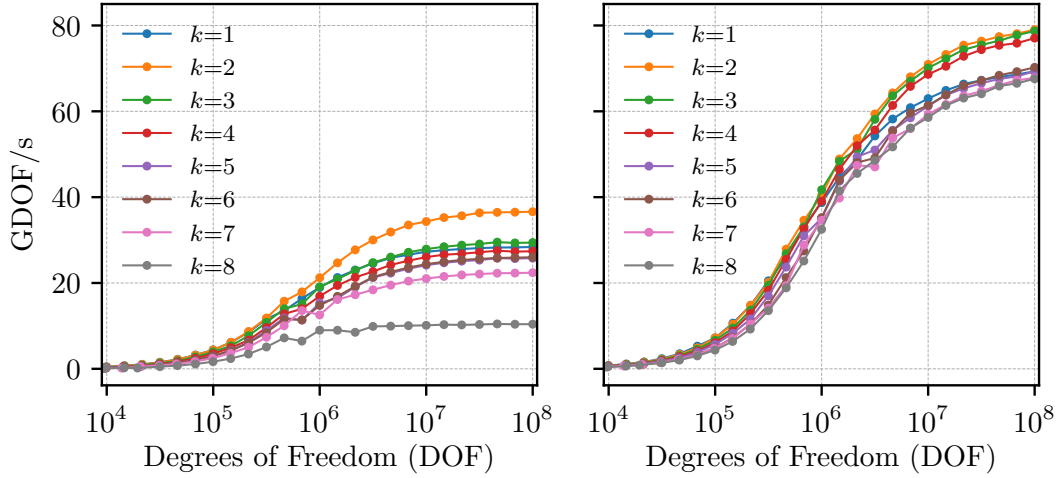}%
    }
    \caption{FP32 performance comparison of the batched vector Raviart--Thomas mass operator ($\mathbf{M}$) on a single AMD MI250X die (left) and an NVIDIA GH200 (right) as a function of the polynomial degree $k$.}
    \label{fig:AMD_NV_Mass}
\end{figure}

Vector units, i.e., stream processors on AMD and NVIDIA GPUs, are leveraged to execute the compute work of the scalar Laplace operator ($\mathbf{L}$) and the vector Raviart--Thomas mass operator ($\mathbf{M}$) in a benchmark akin to the CEED bakeoff kernel BK3 (but with $k+1$ quadrature points per direction, labeled BK3.5) and the Raviart--Thomas equivalent of BK2~\cite{Fischer20scalability}, respectively.
We follow the workload mapping detailed in
Table~\ref{tab:gpu_mapping_compute}. Figure~\ref{fig:AMD_NV_Mass} presents the 
throughput measured across a wide range of problem sizes on the two architectures, 
showing that after an initial transient phase, the performance reaches a plateau.

To further investigate the measured throughput, we used NVIDIA Nsight Compute to collect performance metrics. Based on these measurements, we present the Roofline model in Figure~\ref{fig:roofline_model} for the vector mass operator ($\mathbf{M}$) with $10^8$ degrees of freedom (DOF) on an NVIDIA GH200, including both FP32 and FP64 precision.
As the polynomial order increases, the arithmetic intensity rises, enabling higher attainable performance within the roofline limits. Moreover, FP32 reduces the number of bytes transferred per DOF by half compared with FP64, thereby doubling the arithmetic intensity and yielding a speedup of up to 1.7$\times$ over FP64. While all data points lie primarily in the memory-bound regime, there remains a small gap to peak bandwidth due to interaction with two other limitations, shared-memory bandwidth and thread utilization.

\definecolor{primary}{HTML}{2d74ca}   
\definecolor{secondary}{HTML}{b82231} 

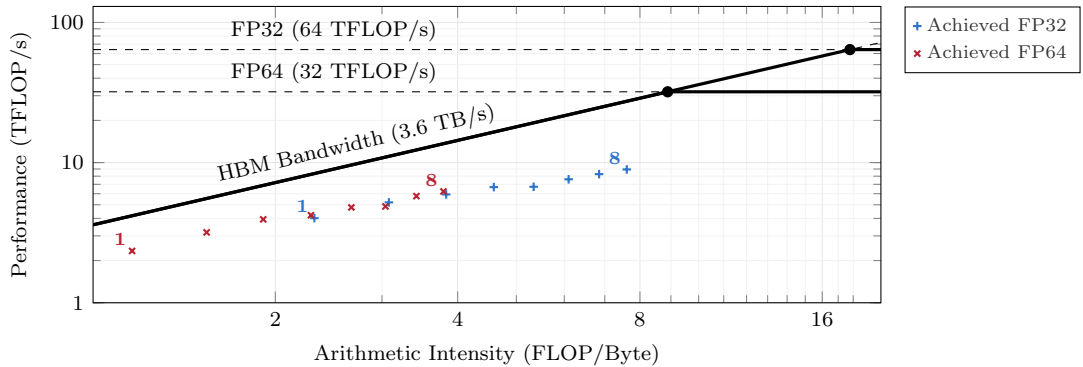
\begin{figure}[htbp]
\centering
\begin{tikzpicture}
\begin{axis}[
    xmode=log,
    ymode=log,
    log basis x={10},
    log basis y={10},
    log ticks with fixed point,
    xtick={2, 4, 8, 16},
    minor xtick={3, 5, 6, 7, 9, 10, 11, 12, 13, 14, 15, 17, 18},
    xlabel={Arithmetic Intensity (FLOP/Byte)},
    ylabel={Performance (TFLOP/s)},
    label style={font=\footnotesize},
    tick label style={font=\footnotesize},
    xmin=1, xmax=20,
    ymin=1, ymax=130,
    grid=both,
    minor grid style={line width=.1pt, draw=gray!10},
    major grid style={line width=.3pt, draw=gray!20},
    width=0.75\linewidth,
    height=5.5cm,
    legend pos=outer north east,
    legend style={font=\scriptsize, fill=white, fill opacity=0.8, draw opacity=1},
    legend image code/.code={
        \draw[#1] plot coordinates {(0.3cm,0cm)};
    }
]

\addplot [domain=1:128, samples=100, dashed, color=black, line width=0.4pt, forget plot] {3.6*x}
    node[pos=0.21, sloped, above, yshift=2pt, inner sep=1pt, font=\footnotesize, color=black] {HBM Bandwidth (3.6 TB/s)};

\addplot [domain=1:8.889, dashed, color=black, line width=0.4pt, forget plot] {32};
\node[above, yshift=2pt, inner sep=1pt, font=\footnotesize, color=black] at (axis cs:2.5, 32) {FP64 (32 TFLOP/s)};

\addplot [domain=1:128, samples=300, color=black, line width=1.2pt, forget plot] {min(3.6*x, 32)};

\addplot [domain=1:17.778, dashed, color=black, line width=0.4pt, forget plot] {64};
\node[above, yshift=2pt, inner sep=1pt, font=\footnotesize, color=black] at (axis cs:2.5, 64) {FP32 (64 TFLOP/s)};

\addplot [domain=1:128, samples=300, color=black, line width=1.2pt, forget plot] {min(3.6*x, 64)};

\node[circle, fill=black, inner sep=1.5pt] at (axis cs:8.889, 32) {};
\node[circle, fill=black, inner sep=1.5pt] at (axis cs:17.778, 64) {};


\addplot[
    only marks,
    mark=+,
    mark options={scale=0.8, line width=0.8pt},
    color=primary
] coordinates {
    (2.32, 4.01375423226353)
    (3.08, 5.20220704985710)
    (3.83, 5.92802334410388)
    (4.59, 6.68714965063510)
    (5.34, 6.72245932632148)
    (6.1, 7.60096969696970)
    (6.85, 8.27403768086817)
    (7.61, 8.93818689453869)
};
\addlegendentry{Achieved FP32}

\node[above left, font=\scriptsize\bfseries, color=primary, inner sep=2pt] at (axis cs:2.32, 4.01375) {1};
\node[above left, font=\scriptsize\bfseries, color=primary, inner sep=2pt] at (axis cs:7.61, 8.93818) {8};

\addplot[
    only marks,
    mark=x,
    mark options={scale=0.8, line width=0.8pt},
    color=secondary
] coordinates {
    (1.16, 2.34367154552442)
    (1.54, 3.17945014257775)
    (1.91, 3.93658322229111)
    (2.29, 4.21337269096355)
    (2.67, 4.79086090915511)
    (3.04, 4.86652481558803)
    (3.42, 5.76405020328087)
    (3.79, 6.22722664308871)
};
\addlegendentry{Achieved FP64}

\node[above left, font=\scriptsize\bfseries, color=secondary, inner sep=2pt] at (axis cs:1.16, 2.34367) {1};
\node[above left, font=\scriptsize\bfseries, color=secondary, inner sep=2pt] at (axis cs:3.79, 6.22722) {8};

\end{axis}
\end{tikzpicture}
\caption{Empirical Roofline Model for the batched vector mass ($\mathbf{M}$) operator performance at a fixed problem size of $10^8$ degrees of freedom (DOF) on NVIDIA GH200. 
Each precision comprises eight data points representing polynomial orders $k = 1$ to $8$, where arithmetic intensity increases with $k$.
}
\label{fig:roofline_model}
\end{figure}

\begin{figure}
	\centering
	\begin{tikzpicture}
		
		\definecolor{primary}{HTML}{5E81AC}   
		\definecolor{secondary}{HTML}{BF616A} 
		
		\begin{axis}[
			name=ax1,
			xlabel={Polynomial order $k$},
			ylabel={GDOF/s},
			grid=major,
			grid style={lightgray!30}, 
			width=0.48\textwidth,       
			height=5.5cm,
			ymin=0, ymax=21,           
			xmin=1, xmax=8,
			xtick distance=1,
			ytick distance=2,
			font=\footnotesize,
			legend pos=north east,
			legend style={draw=none, fill=white, font=\scriptsize, legend cell align={left}} 
			]
			
			\addplot+[color=primary, thick, mark=*, mark options={fill=primary}] coordinates {
				(1, 5.781e-01) (2, 2.650e+00) (3, 6.749e+00) (4, 1.052e+01) (5, 1.300e+01) (6, 1.285e+01) (7, 1.331e+01) (8, 1.017e+01)
			};
			\addlegendentry{FP32-Single}
			
			\addplot+[color=secondary, thick, mark=*, mark options={fill=secondary}] coordinates {
				(1, 5.290e+00) (2, 9.756e+00) (3, 1.455e+01) (4, 1.472e+01) (5, 1.341e+01) (6, 1.301e+01) (7, 1.336e+01) (8, 1.021e+01)
			};
			\addlegendentry{FP32-Batched}
			
		\end{axis}
		
		\begin{axis}[
			name=ax2,
			at={(ax1.outer east)},
			anchor=outer west,
			xshift=0.5cm,
			xlabel={Polynomial order $k$},
			grid=major,
			grid style={lightgray!30},
			width=0.48\textwidth, 
			height=5.5cm,
			ymin=0, ymax=21,     
			xmin=1, xmax=8,
			xtick distance=1,
			ytick distance=2,
			font=\footnotesize,
			legend pos=north east,
			legend style={draw=none, fill=white, font=\scriptsize, legend cell align={left}}
			]
			
			\addplot+[color=primary, thick, mark=*, mark options={fill=primary}] coordinates {
				(1, 4.630e-01) (2, 1.898e+00) (3, 3.789e+00) (4, 5.618e+00) (5, 6.822e+00) (6, 6.861e+00) (7, 4.984e+00) (8, 3.706e+00)
			};
			\addlegendentry{FP64-Single}
			
			\addplot+[color=secondary, thick, mark=*, mark options={fill=secondary}] coordinates {
				(1, 2.524e+00) (2, 5.057e+00) (3, 7.282e+00) (4, 6.434e+00) (5, 6.773e+00) (6, 6.816e+00) (7, 4.990e+00) (8, 3.708e+00)
			};
			\addlegendentry{FP64-Batched}
			
		\end{axis}
		
	\end{tikzpicture}
	\caption{Performance comparison of the Laplace ($\mathbf{L}$) operator evaluation at a fixed problem size of $10^8$ degrees of freedom (DOF) on a
		single AMD MI250X die. For the batched technique, only one element fits in a single thread block for FP32 (polynomial order $k \geq 6$) and FP64 (polynomial order $k \geq 5$).}
	\label{fig:laplace_comparison}
\end{figure}
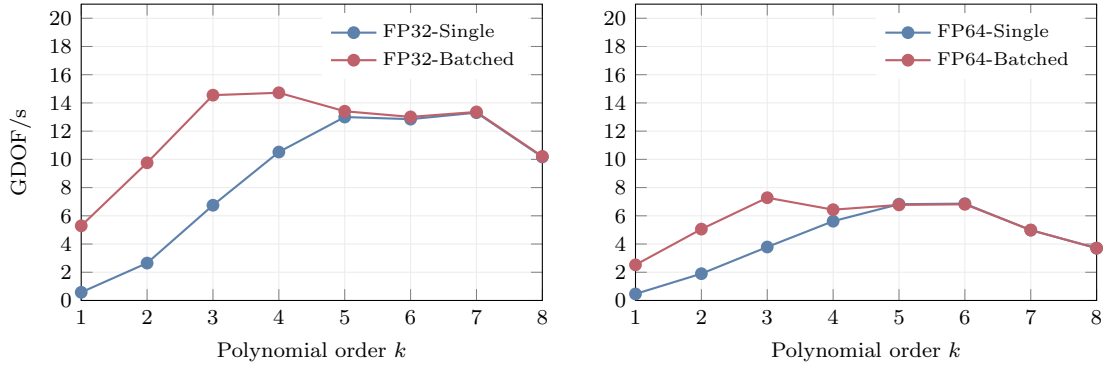

For increased clarity, we present the throughput for different
polynomial degrees in single and double precision in
Figure~\ref{fig:mass_comparison}. The results show that our batching
approach achieves up to a $5\times$ speedup for the Raviart--Thomas mass operator $\mathbf{M}$
over the single-element block counterpart for low
degrees. Figure~\ref{fig:laplace_comparison} shows that similar speedups are
obtained for the scalar Laplace operator $\mathbf{L}$, with batched low-order
elements ultimately outperforming the high-order elements. Note that the
throughput-per-size behaves similarly to the one in Fig.~\ref{fig:AMD_NV_Mass}
and is not shown separately. The figures also illustrate the increase in
throughput for lower-precision arithmetic, primarily driven by the reduced memory access  of a primarily memory-bound step, see the roofline analysis of Fig.~\ref{fig:roofline_model}.

\begin{figure}
    \centering
    \begin{tikzpicture}

        \definecolor{primary}{HTML}{5E81AC}   
        \definecolor{secondary}{HTML}{BF616A} 
        
        \begin{axis}[
            name=ax1,
            xlabel={Polynomial order $k$},
            ylabel={GDOF/s},
            grid=major,
            grid style={lightgray!30}, 
            width=0.48\textwidth,       
            height=5.5cm,
            ymin=0, ymax=52,           
            xmin=1, xmax=8,
            xtick distance=1,
            ytick distance=5,
            font=\footnotesize,
            legend pos=north east,
            legend style={draw=none, fill=white, font=\scriptsize, legend cell align={left}} 
        ]
        
        \addplot+[color=primary, thick, mark=*, mark options={fill=primary}] coordinates {
            (1, 6.925e+00) (2, 1.696e+01) (3, 2.440e+01) (4, 2.611e+01) (5, 2.564e+01) (6, 2.585e+01) (7, 2.226e+01) (8, 1.030e+01)
        };
        \addlegendentry{FP32-Single}
        
        \addplot+[color=secondary, thick, mark=*, mark options={fill=secondary}] coordinates {
            (1, 2.838e+01) (2, 3.661e+01) (3, 2.942e+01) (4, 2.740e+01) (5, 2.579e+01) (6, 2.599e+01) (7, 2.240e+01) (8, 1.041e+01)
        };
        \addlegendentry{FP32-Batched}
        
        \end{axis}

        \begin{axis}[
            name=ax2,
            at={(ax1.outer east)},
            anchor=outer west,
            xshift=0.5cm,
            xlabel={Polynomial order $k$},
            grid=major,
            grid style={lightgray!30},
            width=0.48\textwidth, 
            height=5.5cm,
            ymin=0, ymax=52,     
            xmin=1, xmax=8,
            xtick distance=1,
            ytick distance=5,
            font=\footnotesize,
            legend pos=north east,
            legend style={draw=none, fill=white, font=\scriptsize, legend cell align={left}}
        ]
        
        \addplot+[color=primary, thick, mark=*, mark options={fill=primary}] coordinates {
            (1, 5.199e+00) (2, 9.327e+00) (3, 1.313e+01) (4, 1.428e+01) (5, 1.402e+01) (6, 9.886e+00) (7, 1.136e+01) (8, 5.403e+00)
        };
        \addlegendentry{FP64-Single}
        
        \addplot+[color=secondary, thick, mark=*, mark options={fill=secondary}] coordinates {
            (1, 1.615e+01) (2, 1.821e+01) (3, 1.685e+01) (4, 1.433e+01) (5, 1.406e+01) (6, 9.924e+00) (7, 1.139e+01) (8, 5.438e+00)
        };
        \addlegendentry{FP64-Batched}

        \end{axis}
        
    \end{tikzpicture}
    \caption{Performance comparison of the vector mass ($\mathbf{M}$) operator evaluation at a fixed problem size of $10^8$ degrees of freedom (DOF) on a
single AMD MI250X die. For the batched technique, only one element fits in a single thread block for FP32 (polynomial order $k \geq 5$) and FP64 (polynomial order $k\geq 4$).}
    \label{fig:mass_comparison}
\end{figure}
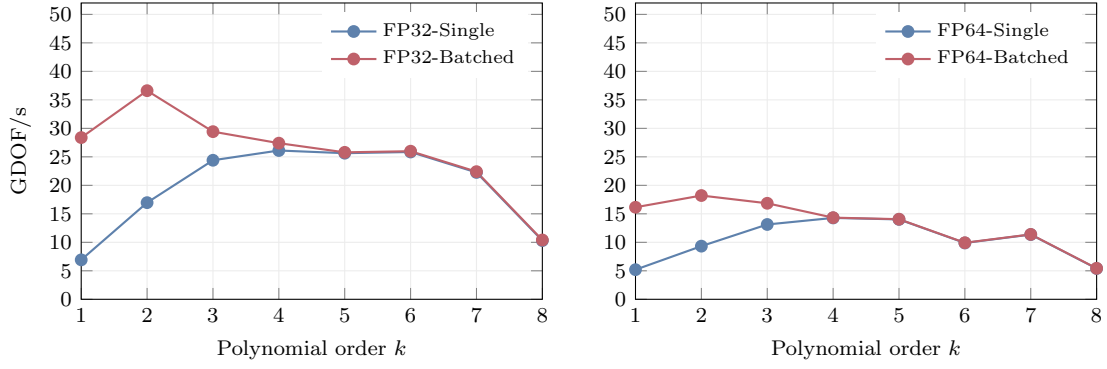

\subsection{Multigrid solver}\label{sec:multigrid}

Solving the pressure Poisson equation~\eqref{eq:pressure_weak} often
constitutes the computational bottleneck in incompressible Navier--Stokes
solvers as the high condition number of the discretized linear
system~\eqref{eq:pressure_lin} requires efficient preconditioning strategies. As
pointed out in Section~\ref{sec:solver_algorithms}, multigrid methods offer
optimal complexity for elliptic problems, and when combined with matrix-free
operator evaluation provide excellent parallel performance, which was
extensively validated on CPUs ~\cite{Arndt20exadg, Kronbichler18performance,
  Munch23efficient}.

In the present work, the pressure Poisson equation is solved by the Conjugate
Gradient (CG) algorithm preconditioned by a multigrid V--cycle. The multigrid
hierarchy of coarser problem representations is constructed by first
polynomial coarsening from the high degree finite element to degree 1, which
is then followed by the global geometric coarsening as described
in~\cite{Munch23efficient}.

For the discontinuous Galerkin discretization of the pressure Poisson
equation~\eqref{eq:pressure_weak}, two aspects of this solver require
particular attention. First, in the matrix-free operator evaluation, face
integration is carried out using the face-centric loop (FCL) approach, with
temporary global storage of the face values and normal derivatives, as
described, e.g., in~\cite{KronbichlerAllalen2018,Kronbichler19fast}. Second,
as shown in~\cite{Fehn20hybrid}, additionally embedding the discontinuous 
operator into the corresponding continuous space at the top of the multigrid
hierarchy leads to better multigrid efficiency and lower iteration counts.

\begin{figure}
	\centering
			\begin{minipage}{0.5\textwidth}
				\begin{center}
	\begin{tikzpicture}[scale=0.6]
				
				\tikzset{
				gridnode/.style={circle, draw, fill=white, inner sep=0pt, minimum size=3pt},
				vnode/.style={circle, draw, fill=white, inner sep=0pt, minimum size=6pt},
				gridplane/.style={draw=black!50, fill=black!5, thick},
				plane_offset/.style={shift={(-30:0.15)}}, 
				edge_label/.style={font=\footnotesize},
				section_label/.style={font=\small},
				transfer_label/.style={font=\footnotesize},
				arrow/.style={-stealth, thick},
				node distance=1.5cm and 2.5cm
			}
			
			
			\node (ph_top_left)  [vnode] at (-5, 7) {};
			\node (ph_top_right) [vnode] at (-1, 7) {};
			\node (ph_mid_left)  [vnode] at (-4, 5) {};
			\node (ph_mid_right) [vnode] at (-2, 5) {};
			\node (ph_bot)       [vnode] at (-3, 3) {};
			
			
			\draw[thick] (ph_top_left) -- (ph_mid_left);
			\draw[thick] (ph_mid_left) -- (ph_bot);
			\draw[thick] (ph_bot)       -- (ph_mid_right);
			\draw[thick] (ph_mid_right) -- (ph_top_right);
			\node [edge_label, anchor=east] at (-5.2, 7) {smoothen};
			\node [edge_label, anchor=west] at (-0.8, 7) {smoothen};
			
			\node [edge_label, anchor=east] at (-4.2, 5) {smoothen};
                        \node [edge_label, anchor=west] at (-1.8, 5) {smoothen};
			\node [edge_label, anchor=west] at (-1.3, 6) {prolongate};
			\node [edge_label, anchor=west] at (-2.3, 4) {prolongate};
			\node [edge_label, anchor=east] at (-4.7, 6) {restrict};
			\node [edge_label, anchor=east] at (-3.7, 4) {restrict};
			\node [edge_label, font=\footnotesize] at ($(ph_bot)+(0,-0.5)$) {coarse solve};
	\end{tikzpicture}
\end{center}
\end{minipage}
	\begin{minipage}{0.49\textwidth}
	\begin{center}
			\begin{tikzpicture}[scale=0.6]
				\centering
				\tikzset{
					gridnode/.style={circle, draw, fill=white, inner sep=0pt, minimum size=3pt},
					vnode/.style={circle, draw, fill=white, inner sep=0pt, minimum size=6pt},
					gridplane/.style={draw=black!50, fill=black!5},
					plane_offset/.style={shift={(-30:0.15)}}, 
					edge_label/.style={font=\footnotesize},
					section_label/.style={font=\small},
					transfer_label/.style={font=\footnotesize},
					arrow/.style={-stealth, thick},
					node distance=1.5cm and 2.5cm
				}
				
				%
				\draw[gridplane] (0,0) rectangle (2,2);
				\foreach \x in {0.5, 1, 1.5}
                                {
                                  \draw (\x, 0) -- (\x, 2);
                                  \draw (0, \x) -- (2, \x);
                                }
				\foreach \x in {0, 0.5, 1, 1.5, 2}
				  \foreach \y in {0, 0.5, 1, 1.5, 2}
				    \node [gridnode] at (\x, \y) {};
				\node (h_top) at (1, 1) {};
				%
				\begin{scope}[shift={(0, -3.5)}]
                                  \draw[gridplane] (0,0) rectangle (2,2);
                                  \draw (1, 0) -- (1, 2);
                                  \draw (0, 1) -- (2, 1);
                                  \foreach \x in {0,1,2}
				    \foreach \y in {0,1,2}
				      \node [gridnode] at (\x, \y) {};
				  \node [gridnode] at (1, 1) {};
				\end{scope}
				\node (h_bot) at (1.5, -3.5) {};
				
				\draw[arrow] (1, -0.2) -- (1, -1.2);
				\node [transfer_label, right=0.2cm] at (1.25, -0.75) {\textit{h}-transfer};
				
					\begin{scope}[shift={(5, 0)}]
				\draw[gridplane] (0,0) rectangle (2,2);
				\foreach \x in {0,1,2} \draw[black!50] (\x,0) -- (\x,2);
				\foreach \y in {0,1,2} \draw[black!50] (0,\y) -- (2,\y);
				\foreach \x in {0, 0.5, 1, 1.5, 2}
				\foreach \y in {0, 0.5, 1, 1.5, 2}
				\node [gridnode] at (\x, \y) {};
				
				\begin{scope}[shift={(0, -3.5)}]
					\draw[gridplane] (0,0) rectangle (2,2);
					\draw[black!50] (1,0) -- (1,2);
					\draw[black!50] (0,1) -- (2,1);
					\foreach \x in {0, 1, 2}
					\foreach \y in {0,1,2}
					\node [gridnode] at (\x, \y) {};
				\end{scope}
				
				\draw[arrow] (1, -0.2) -- (1, -1.2);
				\node [transfer_label, right=0.2cm] at (1.25, -0.75) {\textit{p}-transfer};
					\end{scope}
		\end{tikzpicture}
\end{center}
	\end{minipage}
	\caption{Illustration of the multigrid V--cycle (left) and transfer operators (right).}
	\label{fig:v_cycle}
\end{figure}
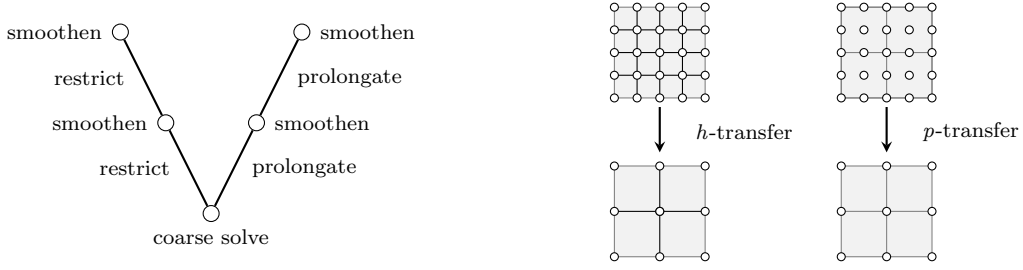

The multigrid preconditioner is formulated as a recursive iterative procedure
illustrated in~Fig.~\ref{fig:v_cycle} with an interplay of three main
components: smoothers, prolongation/restriction transfer operators and a coarse
solver. The choice of these three ingredients incorporating highly efficient
matrix-free computations based on sum factorization is of paramount
importance to achieving the best parallel performance of the solver. In a
matrix-free setting the choice of smoothers is limited, as effective schemes
such as Gauss--Seidel relaxation or incomplete factorizations directly rely on
matrix entry substitutions, and the derivation of efficient matrix-free smoother
is an active research topic~\cite{Brubeck22,Cui2025PatchSmootherGPU,Munch24cache,Phillips25,Stiller17,Sundar15}. Among matrix-vector based smoothers, a polynomial
Chebyshev-accelerated pointwise Jacobi smoother is a good practical choice as,
besides matrix-vector products, it requires only the matrix diagonal which is
precomputed and stored in the solver setup
phase~\cite{ADAMSChebyshevSmoother}. The transfer operators on tensor-product
finite element spaces, both for polynomial and geometric
prolongation/restriction operations, can be effectively evaluated using sum
factorization, which allows us to adopt the highly optimized kernel design as
discussed in the previous subsection. For the coarse solver, the Chebyshev
smoother is employed as a direct solver to achieve accuracy of $10^{-3}$,
comparable to the accuracy of the other multigrid levels.

\input{figures/poisson_mg_solver.tex}

Figure~\ref{fig:mg_solver} shows the performance of the continuous and discontinuous 
multigrid solvers for polynomial degree $k=4$ on a 3D cube with slightly deformed 
elements on NVIDIA GH200 GPUs on the EuroHPC supercomputer JUPITER. The multigrid
hierarchy of coarser representations for this benchmark has been chosen by
polynomial coarsening from degree 4 to degree 2 and to degree~1, followed by
the geometric global coarsening to the coarsest mesh with a single box
element, thus maintaining a constant ratio of degrees of freedom between each
multigrid level of approximately 8. For the discontinuous solver, additional embedding
into the continuous space is carried out at the top level. Two variants of the multigrid
implementation are compared: mixed FP32/FP64 precision, where FP32 is used
for the V--cycle preconditioner and only the operations of the outer CG solver
are computed using FP64, and high precision, where FP64 is used throughout the
entire solver. We note that pure FP32 execution according to Algorithm~\ref{alg:mixed_precision}
is also possible if the application allows for it. A clear benefit
from using mixed precision is observed in terms of throughput.

	\pgfplotstableread{
	gpus  cells     dofs    mv_outer  mv_inner   cg_time  cg_its cg_reduction
	4     4194304 269748225 7.045e-03 4.535e-03 5.073e-01 5      1.478e-02
	8     8388608 538970625 7.256e-03 4.736e-03 5.667e-01 5      1.444e-02
	16    16777216 1076890625 8.113e-03 5.235e-03 6.586e-01 5      1.523e-02
	32    33554432 2152730625 8.790e-03 5.693e-03 7.459e-01 5      1.510e-02
	64    67108864 4303361025 9.305e-03 6.017e-03 8.697e-01 5      1.535e-02
	128   134217728 8602523649 9.601e-03 6.153e-03 1.095e+00 6      1.540e-02
	256   268435456 17200848897 9.702e-03 6.128e-03 1.179e+00 6      1.577e-02
	512   536870912 34393303041 9.806e-03 6.244e-03 1.328e+00 6      1.672e-02
}\tableMGJupiterWeakFour
\pgfplotstableread{
	gpus  cells     dofs    mv_outer  mv_inner   cg_time  cg_its cg_reduction
	4     2097152 263374721 6.017e-03 4.130e-03 5.374e-01 6      2.152e-02
	8     4194304 526338561 6.342e-03 4.320e-03 6.096e-01 6      2.109e-02
	16    8388608 1051856001 7.208e-03 4.791e-03 7.134e-01 6      2.103e-02
	32    16777216 2102071041 8.280e-03 5.412e-03 8.369e-01 6      2.315e-02
	64    33554432 4202501121 8.881e-03 5.814e-03 1.013e+00 6      2.260e-02
	128   67108864 8401721601 9.076e-03 5.929e-03 1.042e+00 6      2.250e-02
	256   134217728 16796884481 9.165e-03 5.978e-03 1.115e+00 6      2.354e-02
	512   268435456 33587210241 9.203e-03 5.988e-03 1.255e+00 6      2.341e-02
}\tableMGJupiterWeakFive
\pgfplotstableread{
	gpus  cells     dofs    mv_outer  mv_inner   cg_time  cg_its cg_reduction
	4     1048576 227673985 5.963e-03 4.037e-03 4.591e-01 5      9.579e-03
	8     2097152 454756609 6.518e-03 4.357e-03 5.336e-01 5      1.075e-02
	16    4194304 908921857 7.276e-03 4.841e-03 6.138e-01 5      1.093e-02
	32    8388608  1816661761 8.345e-03 5.416e-03 7.295e-01 5      1.124e-02
	64    16777216 3630961153 9.320e-03 5.981e-03 8.806e-01 5      1.244e-02
	128   33554432 7259559937 9.846e-03 6.164e-03 9.666e-01 5      1.293e-02
	256   67108864 14514396673 9.675e-03 6.174e-03 9.967e-01 5      1.392e-02
	512   134217728 29019350017 9.799e-03 6.235e-03 1.199e+00 5      1.526e-02
}\tableMGJupiterWeakSix

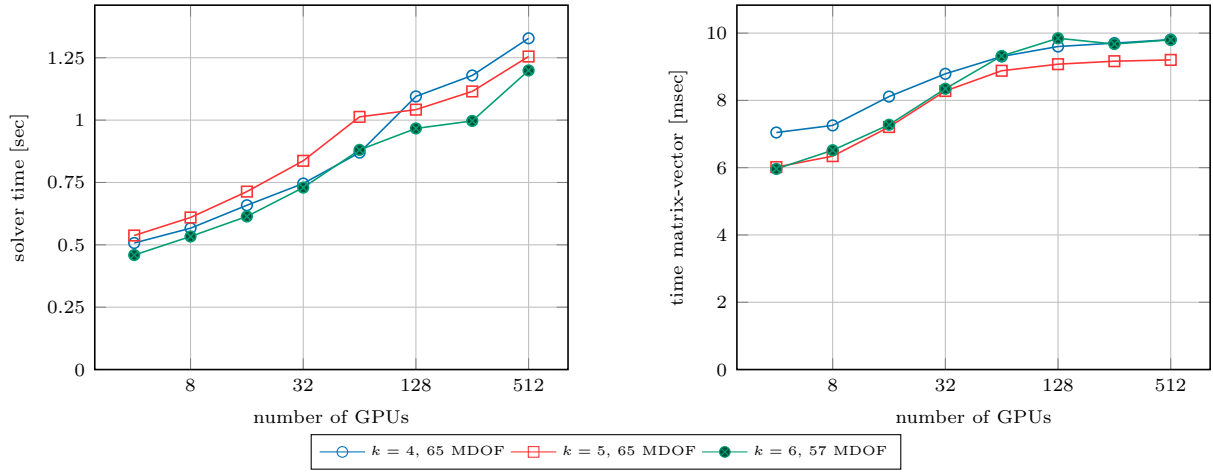
\begin{figure}[htbp]
		\begin{tikzpicture}
		\begin{semilogxaxis}[
			title style={font=\scriptsize},
			tick label style={font=\scriptsize},
			label style={font=\scriptsize},
			legend style={font=\tiny},
			width=0.49\columnwidth,
			height=0.4\columnwidth,
			ylabel={solver time [sec]},
			xlabel={number of GPUs},
			legend columns = 3,
			legend to name=legend:weakmv,
			legend cell align={left},
			cycle list name=colorGPL,
			grid,
			semithick,
			ymin=0,
			ytick={0,0.25,0.5,0.75,1,1.25},
			xtick={2,8,32,128},
			xticklabels={8,32,128,512},
			]
			\addplot table[x expr={\thisrowno{0}/4}, y expr={\thisrowno{5}}] {\tableMGJupiterWeakFour};
			\addlegendentry{$k=4$, 65 MDOF};
			\addplot table[x expr={\thisrowno{0}/4}, y expr={\thisrowno{5}}] {\tableMGJupiterWeakFive};
			\addlegendentry{$k=5$, 65 MDOF};
			\addplot table[x expr={\thisrowno{0}/4}, y expr={\thisrowno{5}}] {\tableMGJupiterWeakSix};
			\addlegendentry{$k=6$, 57 MDOF};
		\end{semilogxaxis}
				\end{tikzpicture}
		\hfill
		\begin{tikzpicture}
			\begin{semilogxaxis}[
				title style={font=\scriptsize},
				tick label style={font=\scriptsize},
				label style={font=\scriptsize},
				legend style={font=\tiny},
				width=0.49\columnwidth,
				height=0.4\columnwidth,
				ylabel={time matrix-vector [msec]},
				xlabel={number of GPUs},
				legend columns = 3,
				legend cell align={left},
				cycle list name=colorGPL,
				grid,
				semithick,
				ymin=0,
				ytick={0,2,4,6,8,10,12},
				xtick={2,8,32,128},
				xticklabels={8,32,128,512},
				]
				\addplot table[x expr={\thisrowno{0}/4}, y expr={\thisrowno{3}*1e3}] {\tableMGJupiterWeakFour};
				\addplot table[x expr={\thisrowno{0}/4}, y expr={\thisrowno{3}*1e3}] {\tableMGJupiterWeakFive};
				\addplot table[x expr={\thisrowno{0}/4}, y expr={\thisrowno{3}*1e3}] {\tableMGJupiterWeakSix};
			\end{semilogxaxis}
	\end{tikzpicture}
		\\ \strut\hfill\pgfplotslegendfromname{legend:weakmv}\hfill\strut
		\caption{Weak scaling of the multigrid solver (left) and matrix-vector evaluation (right) on NVIDIA GH200 GPUs (JUPITER)}
		\label{fig:mg-weak-scaling}
\end{figure}

The weak scaling of the continuous multigrid solver and the matrix-vector product for
polynomial degrees $k=4,5, 6$ on up to 128 nodes/512 NVIDIA GH200 GPUs of
JUPITER is presented in Fig.~\ref{fig:mg-weak-scaling}. The results show a
very good scaling of matrix--vector product, with a parallel efficiency around
65\%. However, the multigrid efficiency is not as good with run time
increasing by a factor of 2.5 when increasing computational resources by a
factor of 128, despite the algorithmic optimality of the method. This can be
traced back to a poorer behavior on the coarser multigrid levels. Indeed,
initial investigation of the communication patterns has shown that around 60\%
of the total solver time is spent on communication and at least half of the
multigrid levels with low DOF counts are almost completely
latency-bound. However, for the discontinuous solver a noticeable plateau is observed in Figure~\ref{fig:mg_solver}, which suggests better scalability properties. This reflects the fact that the finest level more strongly dominates the compute work ($\sim$84\% discontinuous and continuous) than in the continuous case ($\sim$57\%).  Several research
directions to address the communication bottlenecks of the continuous solver and the 
weak scaling are currently being evaluated.

Finally, in Table~\ref{tab:helm_troughput} we report the throughput of the cell-only evaluation and 
full evaluation with face integrals of the Helmholtz operator resulting from the discretization of
the momentum equation~\eqref{eq:ns_momentum} with Raviart--Thomas finite
elements on four NVIDIA GH200 (JUPITER) GPUs. Compared to the mass
operator in Fig.~\ref{fig:mass_comparison}, we observe a lower throughput of the cell-only operator, which illustrates the cost of the additional contributions in the viscous term that necessitate higher shared memory usage. Additionally, the throughput of the full operator with face integrals is from 2.5 to 3 times lower than cell-only evaluation, which reflects the extra work and memory access necessary for face integrals
and three more kernel launches than in the cell-only evaluation due to FCL approach 
adopted here. Alternative approaches to face integration, including cell-centric loop (CCL) and
their performance assessment on GPUs, are subjects of future work.

\begin{table}
	\caption{Throughput in GDOF / [sec $\times$ GPU] for the Helmholtz operator discretized 
  with Raviart--Thomas finite elements on 4 NVIDIA GH200 (JUPITER) GPUs.}
	\label{tab:helm_troughput}
	\small
        \centering
	\begin{tabular}{llcccc}
		\hline
	& 	&\multicolumn{2}{c}   {\textbf{cell only}} & 	\multicolumn{2}{c}  {\textbf{cell + face}}  \\
	&	& FP64  & FP32  & FP64  & FP32 \\
		\hline
		$k=4$ & $787\,\text{MDOF}$ & 13.94 & 21.67 & 4.67 & 7.42 \\
		$k=5$ & $681\,\text{MDOF}$ & 13.05 & 24.52 & 5.13 & 8.55 \\
		\hline
	\end{tabular}
\end{table}

\section{Evaluation of mixed-precision solution algorithm}\label{sec:mixed_precision}

\begin{figure}
  \centering
\includegraphics[height=4cm]{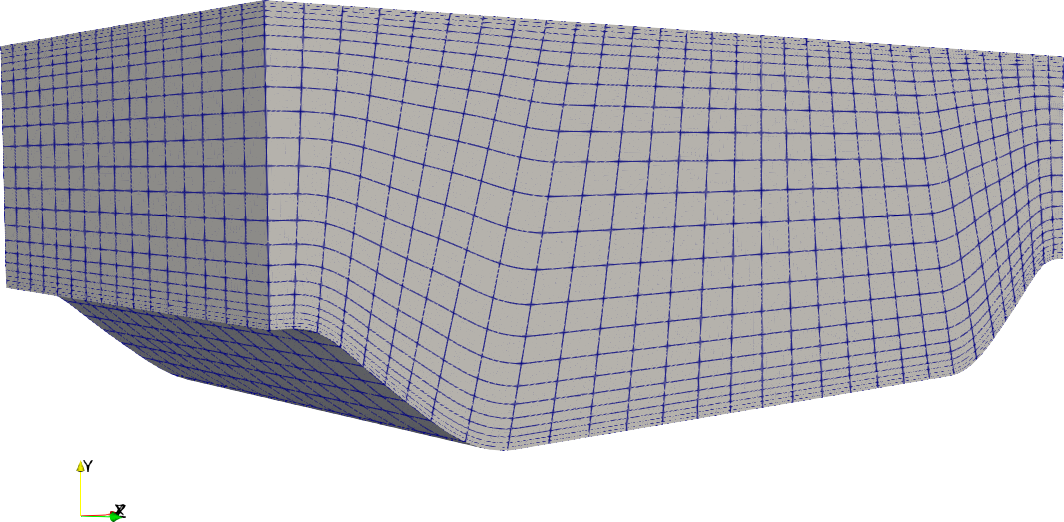}
\quad
\includegraphics[height=4cm]{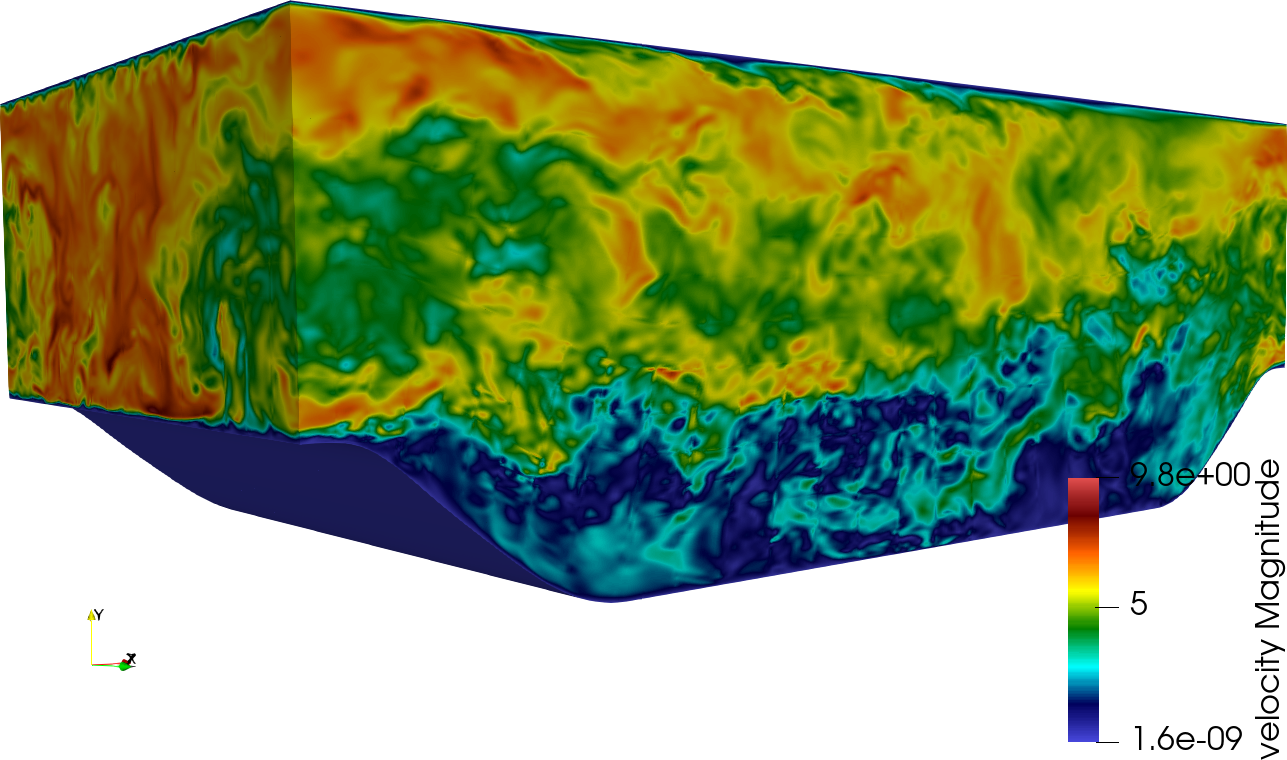}
\caption{Visualization of discretization of the domain into elements (left)
  and the instantaneous velocity field for polynomial degree $k=7$ (right) at
  $t=11\, t_{\text{flow-through}}$ used for testing the mixed-precision
  CFD algorithms.}
\label{fig:hill}
\end{figure}

We evaluate the proposed mixed-precision solver from Algorithm~\ref{alg:mixed_precision} on a real-world CFD test case, using the
flow over a periodic hill at a Reynolds number of $\text{Re}=5600$, following the description
in~\cite{Krank18direct}. A mesh consisting of $64\times 48\times 48$
elements with polynomial degree $k=6$ (tangential degree in RT and pressure
degree), giving 152 million velocity unknowns and 51 million pressure unknowns, is
used for testing. The mesh elements are refined towards the boundary in a
$\tanh(2y) / \tanh(2)$ fashion. A lower-resolution representation of the mesh with $32\times 24\times 16$ elements
is shown in Fig.~\ref{fig:hill}. BDF-3 time stepping ($J=3$) with $\Delta t = 0.32 \frac{h}{(k+1)^{1.5} \|\mathbf u\|_\infty}$ is used. We start the solver from a restart snapshot
at time $t=60\, t_{\text{flow-through}}$ and run for one flow-through time
unit until $t=61\, t_{\text{flow-through}}$, needing around 27,400 time steps. This time span gives a representative statistical impression of the solver behavior, albeit the flow statistics typically need 40--80 flow-through times~\cite{Krank18direct}. The multigrid solver uses a
combination of polynomial and geometric coarsening as described above, with
a crucial additional ingredient being a geometric semi-coarsening in wall-normal
direction at the finest levels to ensure optimality of the
hierarchy~\cite{Trottenberg01multigrid}.

The verification of tolerances as well as the accuracy of mixed-precision
linear solvers from Algorithm~\ref{alg:mixed_precision} can be derived from the accuracy of the initial-guess
projection: If the linear solver accuracy is insufficient, a loss in smoothness is
observed, which leads to a decrease in the projection quality. For each CFD
case, we first monitor the average of the achieved residual reduction by
projection of Alg.~\ref{alg:projective} in (HP) with tight tolerances over 50 time steps. We then work with
the proposed (LP) solvers unless more than 5 subsequent time steps are
$10\times$ above the average (HP) residual reduction.  With this metric,
accurate flow statistics~\cite{Krank18direct} were observed, including other turbulent simulations like the 3D Taylor--Green vortex or flows past cylinders and spheres. For the present case, a
convergence tolerance of $\|b- A x\|_2$ requesting $10^{-8} \|b\|_2$ in the
Euclidean norm for the momentum equations and $10^{-5} \|b\|_2$ for the
pressure Poisson equation was found. To allow finer resolutions with higher
accuracy demands, we select a stricter tolerance of $10^{-9} \|b\|_2$ for the
momentum equation and $10^{-6} \|b\|_2$ for the pressure Poisson equation in
the following.

Our approach is beneficial for all hardware types, including accelerators and classical CPU hardware. Table \ref{tab:initial_guess} reports the costs of projection in comparison to
the actual solve time, computed on four nodes of the CPU cluster
Fritz,\footnote{\url{https://doc.nhr.fau.de/clusters/fritz/}, retrieved on
  June 1, 2026.} each consisting of $2\times 36$ cores of Intel Xeon Platinum
8360Y with $16$ DDR-3200 memory channels. The cost for projection,
including the necessary matrix-vector product with the mass matrix $\mathbf{M}$ for $U^{n+1}_0$, are of similar
magnitude as the cost for the computation of the right-hand side. The times reported in the table do not include the cost for
evaluating the convective terms in~\eqref{eq:ns_pressure}--\eqref{eq:ns_momentum}, which are computed in a fused loop to amortize the
expensive over-integration on $11^3$ Gaussian quadrature points with
Piola-transformed velocity fields to ensure full robustness, compared to $8^3$
points for the left-hand side terms in~\eqref{eq:velocity_weak}, and reported in Table~\ref{tab:solver_mixed_precision}.

\begin{table}
  \centering
  \caption{Compute times and quality metrics of projection step for turbulent
    flow over a periodic hill with 202 million unknowns on 4 compute nodes, averaged over 27,400
    time steps. Compute times and iteration counts are reported as arithmetic means, residual
    reductions as geometric means.}
\label{tab:initial_guess}
\small
\begin{tabular}{lcccccc}
    \hline
     & compute rhs & project & solve & residual reduction & residual reduction & iterations\\
     & time  & time  & time  & by projection & by solver & solve\\
    \hline
    pressure~\eqref{eq:pressure_lin} & 3.4 ms & 3.6 ms & 92.4 ms & $1.39\cdot 10^4$ & $6.03\cdot 10^2$ & 2.02 \\
    momentum~\eqref{eq:velocity_lin} & 12.1 ms & 9.8 ms & 101 ms & $1.79\cdot 10^6$ & $8.41\cdot 10^2$ & 6.22 \\
    \hline
\end{tabular}
\end{table}

\begin{table}
  \caption{Overall run times and average iteration count for turbulent flow on
    periodic hill with 202m unknowns with 27,400 time steps on 4 nodes of the CPU system Fritz. The mixed FP64/FP32 case executes Algorithm~\ref{alg:mixed_precision} with steps performed in the respective precision.}
\label{tab:solver_mixed_precision}
\small
\centering
\begin{tabular}{lcrrcrrc}
  \hline
  & \multicolumn{3}{c}{\textbf{pressure step~\eqref{eq:pressure_lin}}}
  & \multicolumn{3}{c}{\textbf{momentum step~\eqref{eq:velocity_lin}}}
  & \textbf{eval. convection}
  \\
  & number of & time & time  &  number of & time & time  &  time\\
  & iterations & FP32 & FP64  &  iterations & FP32 & FP64  &  FP64\\
  \hline
  all FP64 & 1.93 & --- & 5,350 s & 6.18 & --- & 5,750 s & 912 s \\
  mixed FP64/FP32 & 2.02 & 2,640 s & 81 s & 6.22 & 2,700 s & 672 s & 927 s \\
  \hline
\end{tabular}
\end{table}

The results in Table~\ref{tab:solver_mixed_precision} show a run-time
reduction of $49\%$ for the pressure Poisson and by $41\%$ for momentum solver
steps when performing the proposed mixed-precision approach with the bulk of
work done in FP32, compared to running all steps in the higher precision
FP64. Note that the FP64 proportion of the momentum solver is higher than for
the FP64/FP32 pressure Poisson step due to a cheaper preconditioner.
The iteration counts are only slightly increased by the use of mixed
precision, where roundoff effects cause one more
iteration in a few of the time steps. 
The overall solver time, including all setup and collection of
turbulence statistics similarly to~\cite{Krank18direct}, has decreased from
12,310 seconds to 7,280 seconds with the mixed-precision approach, a reduction
in run time by 41\% (i.e., improving throughput by $1.69\times$), while
maintaining equivalent accuracy.

\section{Summary}\label{sec:summary}

We have developed an efficient, matrix-free implementation of a high-order
finite element solver of the incompressible Navier--Stokes equations with
$H(\text{div})$-conforming Raviart--Thomas elements. These elements have excellent
stability properties, but are challenging to implement efficiently due to the
non-standard finite element spaces
and Piola transformations in a non-conforming setting.  We have demonstrated
ingredients for efficient GPU implementations of the main ingredients, namely
the matrix-free operator evaluation for Raviart--Thomas and Poisson operators
and multigrid preconditioning.  Since application requirements range from
lower-order elements with polynomial degrees $k=2,3$, which are more flexible
for complicated geometries and provide more dissipation in highly
under-resolved scenarios, to high-order cases for high-fidelity large-eddy
simulation or scale-resolving direct numerical simulation, we have
developed GPU tuning that efficiently handles the whole spectrum of degrees.
We have further demonstrated that the use of lower precision can leverage the
lower memory transfer and additional arithmetic capabilities, illustrated by
an end-to-end performance improvement of almost $1.7\times$ for a real-world
case of the flow over a periodic hill with boundary-layer adapted mesh.
Finally, we have outlined the use of matrix units (tensor cores) for the
sum-factorization steps with Raviart--Thomas elements.

While the achieved performance allows us to push the limits of scale-resolving
flow simulations, further improvements are conceivable. The
most important step is the analysis of even lower precisions than FP32: While
the proposed tensor-core algorithm can work with the non-IEEE \texttt{tf32}
format using a 10-bit mantissa rather than 23 bits in FP32, a full use in the
solver has not been presented, as no clear speedup over FP32 could be observed
yet.  We plan to further investigate algorithm variants with lower memory
access and compensated summation~\cite{cui24acceleration},
which might be able to overcome this limit.
Furthermore, it is conceivable to utilize lower precision also in selected
parts of right-hand side evaluation by techniques such as~\cite{Sundriyal26customized} and suitable error control.

\section*{Acknowledgements}
  The authors acknowledge discussions with Shubham K.~Goswami, Katharina
  Kormann, Natalia Nebulishvili, Richard Schussnig, Dominik Still, Carsten Uphoff and
  partners within deal.II.

  This work was funded by the European Research Council (ERC) under the
  European Union's Horizon 2020 research and innovation programme (call
  HORIZON-EUROHPC-JU-2023-COE-03, grant agreement No. 101172493 ``dealii-X'')
  in conjunction with German national co-funding by the Federal Ministry of
  Research, Technology and Space (BMFTR) under the grant agreement numbers
  16HPC112K and 16HPC113.  The authors gratefully acknowledge the European
  High-Performance Computing Joint Undertaking (EuroHPC JU) for awarding this
  project access to the European exascale supercomputer JUPITER (Booster
  module) hosted at the Jülich Supercomputing Centre (JSC) under the project
  id dealii-X@JUPITER, and for access to the EuroHPC supercomputer LUMI, hosted by
  CSC (Finland) and the LUMI consortium through a EuroHPC Regular Access call.
  The authors gratefully acknowledge the scientific support and HPC resources
  provided by the Erlangen National High Performance Computing Center
  (NHR@FAU) of the Friedrich-Alexander-Universität Erlangen-Nürnberg
  (FAU). NHR funding is provided by federal and Bavarian state
  authorities. NHR@FAU hardware is partially funded by the German Research
  Foundation (DFG) -- 440719683.

  \bibliographystyle{plainurl}
  {\footnotesize
    \bibliography{references}

@article{Abdelfattah21survey,
  title = {A survey of numerical linear algebra methods utilizing mixed-precision arithmetic},
  volume = {35},
  DOI = {10.1177/10943420211003313},
  number = {4},
  journal = {Int. J. High Perform. Comput. Appl.},
  author = {Abdelfattah,  Ahmad and Anzt,  Hartwig and Boman,  Erik G. and Carson,  Erin and Cojean,  Terry and Dongarra,  Jack and Fox,  Alyson and Gates,  Mark and Higham,  Nicholas J. and Li,  Xiaoye S and Loe,  Jennifer and Luszczek,  Piotr and Pranesh,  Srikara and Rajamanickam,  Siva and Ribizel,  Tobias and Smith,  Barry F. and Swirydowicz,  Kasia and Thomas,  Stephen and Tomov,  Stanimire and Tsai,  Yaohung M. and Yang,  Ulrike Meier},
  year = {2021},
  pages = {344--369}
}

@article{Abdelfattah21gpu,
  title = {{GPU} algorithms for Efficient Exascale Discretizations},
  volume = {108},
  DOI = {10.1016/j.parco.2021.102841},
  journal = {Parallel Comput.},
  author = {Abdelfattah,  Ahmad and Barra,  Valeria and Beams,  Natalie and Bleile,  Ryan and Brown,  Jed and Camier,  Jean-Sylvain and Carson,  Robert and Chalmers,  Noel and Dobrev,  Veselin and Dudouit,  Yohann and Fischer,  Paul and Karakus,  Ali and Kerkemeier,  Stefan and Kolev,  Tzanio and Lan,  Yu-Hsiang and Merzari,  Elia and Min,  Misun and Phillips,  Malachi and Rathnayake,  Thilina and Rieben,  Robert and Stitt,  Thomas and Tomboulides,  Ananias and Tomov,  Stanimire and Tomov,  Vladimir and Vargas,  Arturo and Warburton,  Tim and Weiss,  Kenneth},
  year = {2021},
  pages = {102841}
}

@article{Abdelfattah21batched,
  title = {A Set of Batched Basic Linear Algebra Subprograms and {LAPACK} Routines},
  volume = {47},
  DOI = {10.1145/3431921},
  number = {3},
  journal = {ACM Transactions on Mathematical Software},
  publisher = {Association for Computing Machinery (ACM)},
  author = {Abdelfattah,  Ahmad and Costa,  Timothy and Dongarra,  Jack and Gates,  Mark and Haidar,  Azzam and Hammarling,  Sven and Higham,  Nicholas J. and Kurzak,  Jakub and Luszczek,  Piotr and Tomov,  Stanimire and Zounon,  Mawussi},
  year = {2021},
  pages = {1–23}
}

@article{Aliaga22compressed,
  title = {Compressed basis {GMRES} on high-performance graphics processing units},
  volume = {37},
  DOI = {10.1177/10943420221115140},
  number = {2},
  journal = {The International Journal of High Performance Computing Applications},
  author = {Aliaga,  José I. and Anzt,  Hartwig and Gr\"{u}tzmacher,  Thomas and Quintana-Ortí,  Enrique S. and Tomás,  Andrés E.},
  year = {2022},
  pages = {82--100}
}

@article{Arnold02unified,
author = {Arnold, Douglas N. and Brezzi, Franco and Cockburn, Bernardo and Marini, L. Donatella},
title = {Unified analysis of discontinuous {G}alerkin methods for elliptic problems},
journal = {SIAM J. Numer. Anal.},
volume = 39,
year = 2002,
pages = {1749--1779},
doi   = {10.1137/S0036142901384162}
}

@article{Austin21initial,
  title = {Initial Guesses for Sequences of Linear Systems in a {GPU}-Accelerated Incompressible Flow Solver},
  volume = {43},
  DOI = {10.1137/20m1368677},
  number = {4},
  journal = {SIAM J. Sci. Comput.},
  publisher = {Society for Industrial & Applied Mathematics (SIAM)},
  author = {Austin,  Anthony P. and Chalmers,  Noel and Warburton,  Tim},
  year = {2021},
  pages = {C259--C289}
}

@article{dealii97,
  title = {The deal.{II} library,  version 9.7},
  volume = {33},
  DOI = {10.1515/jnma-2025-0115},
  number = {4},
  journal = {J. Numer. Math.},
  publisher = {Walter de Gruyter GmbH},
  author = {Arndt,  Daniel and Bangerth,  Wolfgang and Bergbauer,  Maximilian and Blais,  Bruno and Fehling,  Marc and Gassm\"{o}ller,  Rene and Heister,  Timo and Heltai,  Luca and Kronbichler,  Martin and Maier,  Matthias and Munch,  Peter and Scheuerman,  Sam and Turcksin,  Bruno and Uzunbajakau,  Siarhei and Wells,  David and Wichrowski,  Michał},
  year = {2025},
  pages = {403--415}
}

@article{dealiigeneric,
  author = {Daniel Arndt and Wolfgang Bangerth and Denis Davydov and Timo
    Heister and Luca Heltai and Martin Kronbichler and Matthias Maier and 
      Jean-Paul Pelteret and Bruno Turcksin and David Wells},
  title = {The deal.{{II}} finite element library: design, features, and insights},
  doi = {10.1016/j.camwa.2020.02.022},
  journal = {Comput. Math. Appl.},
  volume = {81},
  pages = {407--422},
  year = {2021},
}

@InProceedings{Arndt20exadg,
title     = {{ExaDG}: High-Order Discontinuous {G}alerkin for the Exa-Scale},
DOI = {10.1007/978-3-030-47956-5\_8},
author    = {Arndt, Daniel and Fehn, Niklas and Kanschat, Guido and Kormann, Katharina  
             and Kronbichler, Martin and Munch, Peter and Wall, Wolfgan A. and Witte, Julius}, 
editor    = {Bungartz, H.-J. and Reiz, S. and Uekermann, B.
             and Neumann, P. and Nagel, W.E.},
booktitle = {Software for Exascale Computing -- SPPEXA 2016--2019},
year      = {2020},
publisher = {Springer International Publishing},
address   = {Cham},
pages     = {189--224}
}

@article{Benzi26,
  title = {Scalable augmented {L}agrangian preconditioners for fictitious domain problems},
  volume = {450},
  DOI = {10.1016/j.cma.2025.118522},
  journal = {Computer Methods in Applied Mechanics and Engineering},
  author = {Benzi,  Michele and Feder,  Marco and Heltai,  Luca and Mugnaioni,  Federica},
  year = {2026},
  pages = {118522}
}

@book{Boffi13mixed,
  title = {Mixed Finite Element Methods and Applications},
  ISBN = {9783642365195},
  DOI = {10.1007/978-3-642-36519-5},
  journal = {Springer Series in Computational Mathematics},
  publisher = {Springer Berlin Heidelberg},
  author = {Boffi,  Daniele and Brezzi,  Franco and Fortin,  Michel},
  year = {2013},
  address = {Heidelberg}
}

@article{Brubeck22,
  title = {A Scalable and Robust Vertex-Star Relaxation for High-Order {FEM}},
  volume = {44},
  DOI = {10.1137/21m1444187},
  number = {5},
  journal = {SIAM Journal on Scientific Computing},
  author = {Brubeck,  Pablo D. and Farrell,  Patrick E.},
  year = {2022},
  pages = {A2991--A3017}
}

@article{Clevenger21flexible,
  doi = {10.1145/3425193},
  year = {2021},
  volume = {47},
  number = {1},
  pages = {7/1--27},
  author = {Thomas C. Clevenger and Timo Heister and Guido Kanschat and Martin Kronbichler},
  title = {A Flexible,  Parallel,  Adaptive Geometric Multigrid Method for {FEM}},
  journal = {{ACM} Transactions on Mathematical Software}
}

@article{Creff25,
  title = {Preconditioning of the generalized {S}tokes problem arising from the approximation of the time-dependent {N}avier--{S}tokes equations},
  volume = {191},
  DOI = {10.1016/j.camwa.2025.05.001},
  journal = {Comput. Math. Appl.},
  publisher = {Elsevier BV},
  author = {Creff,  Melvin and Guermond,  Jean-Luc},
  year = {2025},
  pages = {255--274}
}

@article{cui24acceleration,
      title={Acceleration of Tensor-Product Operations with {T}ensor {C}ores}, 
      author={Cu Cui},
      year={2024},
      journal={ACM Trans. Parallel Comput.},
      volume={11},
      number={4},
      pages={15:1--24},
      doi={10.1145/3695466}
}

@incollection{Delorme24,
  title = {A novel mixed precision defect correction solver for heterogeneous computing},
  DOI = {10.1007/978-3-031-73716-9_11},
  booktitle = {High Performance Computing. ISC High Performance 2024 International Workshops},
  publisher = {Springer Nature Switzerland},
  author = {Delorme,  Yann T. and Wasserman,  Mark and Zameret,  Alon and Ding,  Zhaohui},
  editor = {Weiland, M. and Neuwirth, S. and Kruse, C. and Weinzierl, T.},
  year = {2024},
  address = {Cham},
  pages = {154--168}
}

@book{Deville02high,
  title={High-order methods for incompressible fluid flow},
  author={Deville, Michel O. and Fischer, Paul F. and Mund, Ernest H.},
  volume={9},
  year={2002},
  publisher={Cambridge University Press},
  address={Cambridge}
}

@article{Fehn19high,
   title={High-order {DG} solvers for under-resolved turbulent incompressible flows: A comparison of $ {L}^{2}$ and $ {H} $(div) methods},
   author={Niklas Fehn and Martin Kronbichler and Christoph Lehrenfeld and
                  Gert Lube and Philipp W. Schroeder},
   journal={Int. J. Numer. Methods Fluids},
   volume = {91},
   number = 11,
   pages = {533--556},
   year={2019},
   doi={10.1002/fld.4763}
 }

@article{Fehn20hybrid,
  title={Hybrid multigrid methods for high-order discontinuous {G}alerkin discretizations},
  author={Niklas Fehn and Peter Munch and Wolfgang A. Wall and Martin Kronbichler},
  journal={Journal of Computational Physics},
  year={2020},
  volume = {415},
  pages = {109538},
  doi = {10.1016/j.jcp.2020.109538}
}

@article{Fischer98projection,
  title = {Projection techniques for iterative solution of {$Ax=b$} with successive right-hand sides},
  volume = {163},
  DOI = {10.1016/s0045-7825(98)00012-7},
  number = {1-4},
  journal = {Comput. Meth. Appl. Mech. Engrg.},
  publisher = {Elsevier BV},
  author = {Fischer,  Paul F.},
  year = {1998},
  pages = {193--204}
}

@article{Fischer22nekrs,
  title = {{NekRS},  a {GPU}-accelerated spectral element {N}avier–{S}tokes solver},
  volume = {114},
  DOI = {10.1016/j.parco.2022.102982},
  journal = {Parallel Comput.},
  author = {Fischer,  Paul and Kerkemeier,  Stefan and Min,  Misun and Lan,  Yu-Hsiang and Phillips,  Malachi and Rathnayake, Thilina and Merzari, Elia and Tomboulides,  Ananias and Karakus,  Ali and Chalmers,  Noel and Warburton,  Tim},
  year = {2022},
  pages = {102982}
}

@article{Fischer20scalability,
  doi = {10.1177/1094342020915762},
  year = {2020},
  volume = {34},
  number = {5},
  pages = {562--586},
  author = {Paul Fischer and Misun Min and Thilina Rathnayake and Som Dutta and Tzanio Kolev and Veselin Dobrev and Jean-Sylvain Camier and Martin Kronbichler and Tim Warburton and Kasia {\'{S}}wirydowicz and Jed Brown}, 
  title = {Scalability of high-performance {PDE} solvers},
  journal = {Int. J. High Perform. Comput. Appl.}
}

@article{Gholami16fft,
author = {Gholami, Amir and Malhotra, Dhairya and Sundar, Hari and Biros, George}, 
title = {{FFT}, {FMM}, or multigrid? {A} comparative study of state-of-the-art {P}oisson solvers for uniform and nonuniform grids in the unit cube},
journal = {SIAM J. Sci. Comput.},
volume = 38, 
number = 3,
pages = {C280--C306},
year = 2016,
doi = {10.1137/15M1010798}
}

@INPROCEEDINGS{Gropp00performance,
author={Gropp, William D. and Kaushik, Dinesh K. and Keyes, David E. and
Smith, Barry F.},
booktitle={SC '00: Proceedings of the 2000 ACM/IEEE Conference on Supercomputing},
title={Performance modeling and tuning of an unstructured mesh {CFD} application},
year={2000},
address={Dallas, TX},
publisher={ACM},
volume={},
number={},
pages={34},
doi={10.1109/SC.2000.10059},
}

@article{Guermond06overview,
  title = {An overview of projection methods for incompressible flows},
  volume = {195},
  DOI = {10.1016/j.cma.2005.10.010},
  number = {44--47},
  journal = {Comput. Meth. Appl. Mech. Engrg.},
  author = {Guermond,  Jean-Luc and Minev,  Peter D. and Shen,  Jie},
  year = {2006},
  pages = {6011--6045}
}

@book{Hesthaven08nodal,
  title={Nodal discontinuous {G}alerkin methods: Algorithms, analysis, and applications},
  author={Hesthaven, Jan S. and Warburton, Tim},
  year={2008},
series = {Texts in Applied Mathematics},
volume = 54,
  doi={10.1007/978-0-387-72067-8},
  publisher={Springer},
  address={Heidelberg}
}

@article{Higham22mixed,
  title = {Mixed precision algorithms in numerical linear algebra},
  volume = {31},
  DOI = {10.1017/s0962492922000022},
  journal = {Acta Numerica},
  publisher = {Cambridge University Press (CUP)},
  author = {Higham,  Nicholas J. and Mary,  Theo},
  year = {2022},
  pages = {347--414}
}

@Article{Kanschat04multilevel,
  title = {Multi-level methods for discontinuous {G}alerkin {FEM} on locally refined meshes},
  author = {Kanschat, Guido},
  journal = {Comput. \& Struct.},
  number = {28},
  pages = {2437--2445},
  volume = {82},
  year = {2004},
  doi = {10.1016/j.compstruc.2004.04.015}
}

@article{Karp26effects,
  title = {Effects of lower floating-point precision on scale-resolving numerical simulations of turbulence},
  volume = {549},
  DOI = {10.1016/j.jcp.2025.114600},
  journal = {Journal of Computational Physics},
  author = {Karp,  Martin and Stanly,  Ronith and Mukha,  Timofey and Galimberti,  Luca and Toosi,  Siavash and Song,  Hang and Dalcin,  Lisandro and Rezaeiravesh,  Saleh and Jansson,  Niclas and Markidis,  Stefano and Parsani,  Matteo and Bose,  Sanjeeb and Lele,  Sanjiva and Schlatter,  Philipp},
  year = {2026},
  pages = {114600}
}

@article{Kashi26mixed,
  title = {Mixed-precision numerics in scientific applications: survey and perspectives},
  volume = {82},
  DOI = {10.1007/s11227-026-08264-4},
  number = {5},
  journal = {The Journal of Supercomputing},
  author = {Kashi,  Aditya and Lu,  Hao and Brewer,  Wesley and Rogers,  David and Matheson,  Michael and Shankar,  Mallikarjun and Wang,  Feiyi},
  year = {2026},
  pages = {287}
}

@article{Kolev21efficient,
  author = {Kolev,  Tzanio and Fischer,  Paul and Min,  Misun and Dongarra,  Jack and Brown,  Jed and Dobrev,  Veselin and Warburton,  Tim and Tomov,  Stanimire and Shephard,  Mark S. and Abdelfattah,  Ahmad and Barra,  Valeria and Beams,  Natalie and Camier,  Jean-Sylvain and Chalmers,  Noel and Dudouit,  Yohann and Karakus,  Ali and Karlin,  Ian and Kerkemeier,  Stefan and Lan,  Yu-Hsiang and Medina,  David and Merzari,  Elia and Obabko,  Aleksandr and Pazner,  Will and Rathnayake,  Thilina and Smith,  Cameron W. and Spies,  Lukas and Swirydowicz,  Kasia and Thompson,  Jeremy and Tomboulides,  Ananias and Tomov,  Vladimir},
doi = {10.1177/10943420211020803},
title ={Efficient exascale discretizations: High-order finite element methods},
journal = {Int. J. High Perform. Comput. Appl.},
volume = {35},
number = {6},
pages = {527--552},
year = {2021}
}

@article{Krank18direct,
  title = {Direct numerical simulation of flow over periodic hills up to $\text{Re}_{H}$=10,595},
  volume = {101},
  DOI = {10.1007/s10494-018-9941-3},
  number = {2},
  journal = {Flow  Turbul. Combust.},
  author = {Krank,  Benjamin and Kronbichler,  Martin and Wall,  Wolfgang A.},
  year = {2018},
  pages = {521–551}
}

@article{Krank17high,
  title = {A high-order semi-explicit discontinuous {G}alerkin solver for 3{D} incompressible flow with application to {DNS} and {LES} of turbulent channel flow},
  volume = {348},
  DOI = {10.1016/j.jcp.2017.07.039},
  journal = {Journal of Computational Physics},
  author = {Krank,  Benjamin and Fehn,  Niklas and Wall,  Wolfgang A. and Kronbichler,  Martin},
  year = {2017},
  pages = {634--659}
}

@article{Kronbichler12generic,
title = {A generic interface for parallel cell-based finite element operator application},
journal = {Comput. Fluids},
volume = {63},
pages = {135--147},
year = {2012},
author = {Kronbichler, Martin and Kormann, Katharina}
}

@article{Kronbichler19fast,
author = {Kronbichler, Martin and Kormann, Katharina},
title = {Fast matrix-free evaluation of discontinuous {G}alerkin finite element operators}, 
year = {2019},
volume = {45},
number = {3},
journal = {ACM Trans. Math. Softw.},
pages = {29/1--40}
}

@article{Kronbichler19gpu, 
title = {Multigrid for matrix-free high-order finite element computations on graphics processors},
author={Kronbichler, Martin and Ljungkvist, Karl},
journal={ACM Trans. Parallel Comput.},
volume={6},
number={1},
pages={2:1--32},
doi={10.1145/3322813},
year={2019}
}

@article{Kronbichler23enhancing,
  title = {Enhancing data locality of the conjugate gradient method for
                  high-order matrix-free finite-element implementations},
  author = {Martin Kronbichler and Dmytro Sashko and Peter Munch},
  journal = {Int. J. High Perform. Comput. Appl.},
  year = {2023},
  volume = {37},
  number = 2,
  pages = {61--81},
  doi = {10.1177/10943420221107880}
}

@article{Kronbichler18performance,
        title={A performance comparison of continuous and discontinuous {G}alerkin methods with fast multigrid solvers},
        author={Kronbichler, Martin and Wall, Wolfgang A.},
        journal={SIAM J. Sci. Comput.},
        volume={40},
        number={5},
        pages={A3423--A3448},
        year={2018},
        doi={10.1137/16M110455X}
}

@article{Liu09open,
        author = {Liu, Jie},
        journal = {J. Comput. Phys.},
        number = {19},
        pages = {7250--7267},
        title = {Open and traction boundary conditions for the incompressible {N}avier--{S}tokes equations},
        volume = {228},
        year = {2009},
        doi = {10.1016/j.jcp.2009.06.021}
}

@article{Lohner04projective,
  title = {Projective prediction of pressure increments},
  volume = {21},
  DOI = {10.1002/cnm.743},
  number = {4},
  journal = {Communications in Numerical Methods in Engineering},
  publisher = {Wiley},
  author = {L\"{o}hner,  Rainald},
  year = {2004},
  pages = {201--207}
}

@article{Luszczek2024batched,
  title = {Batched sparse and mixed-precision linear algebra interface for efficient use of {GPU} hardware accelerators in scientific applications},
  volume = {160},
  DOI = {10.1016/j.future.2024.06.004},
  journal = {Future Generation Computer Systems},
  author = {Luszczek,  Piotr and Abdelfattah,  Ahmad and Anzt,  Hartwig and Suzuki,  Atsushi and Tomov,  Stanimire},
  year = {2024},
  pages = {359--374}
}

@article{Melenk01fully,
  title = {Fully discrete hp-finite elements: fast quadrature},
  volume = {190},
  DOI = {10.1016/s0045-7825(00)00322-4},
  number = {32--33},
  journal = {Comput. Meth. Appl. Mech. Engrg.},
  author = {Melenk,  Jens M. and Gerdes,  Klaus and Schwab,  Christoph},
  year = {2001},
  pages = {4339--4364}
}

@article{Moura15linear,
  title = {Linear dispersion--diffusion analysis and its application to under-resolved turbulence simulations using discontinuous {G}alerkin spectral/hp methods},
  volume = {298},
  DOI = {10.1016/j.jcp.2015.06.020},
  journal = {J. Comput. Phys.},
  publisher = {Elsevier BV},
  author = {Moura,  Rodrigo C. and Sherwin,  Spencer J. and Peiró,  Joaquim},
  year = {2015},
  pages = {695--710}
}

@article{Moura17eddy,
  title = {On the eddy-resolving capability of high-order discontinuous {G}alerkin approaches to implicit {LES} / under-resolved {DNS} of {E}uler turbulence},
  volume = {330},
  DOI = {10.1016/j.jcp.2016.10.056},
  journal = {J. Comput. Phys.},
  publisher = {Elsevier BV},
  author = {Moura,  Rodrigo C. and Mengaldo,  Gianmarco and Peiró,  Joaquim and Sherwin,  Spencer J.},
  year = {2017},
  pages = {615--623}
}

@article{Munch24cache,
  title = {Cache-optimized and low-overhead implementations of additive {S}chwarz methods for high-order {FEM} multigrid computations},
  volume = {38},
  DOI = {10.1177/10943420231217221},
  number = {3},
  journal = {The International Journal of High Performance Computing Applications},
  author = {Munch,  Peter and Kronbichler,  Martin},
  year = {2023},
  pages = {192--209}
}

@article{Munch23efficient,
  title = {Efficient Distributed Matrix-free Multigrid Methods on Locally Refined Meshes for {FEM} Computations},
  volume = {10},
  DOI = {10.1145/3580314},
  number = {1},
  journal = {ACM Trans. Parallel Comput.},
  author = {Munch,  Peter and Heister,  Timo and Prieto Saavedra,  Laura and Kronbichler,  Martin},
  year = {2023},
  pages = {3/1--38}
}

@article{Orszag80spectral,
title = "Spectral methods for problems in complex geometries",
journal = "J. Comput. Phys.",
volume = "37", 
number = "1",
pages = "70--92",
year = "1980", 
doi = "10.1016/0021-9991(80)90005-4",
author = "Orszag, Steven A."
}

@article{Pazner23gpu,
  title = {End-to-end {GPU} acceleration of low-order-refined preconditioning for high-order finite element discretizations},
  volume = {37},
  DOI = {10.1177/10943420231175462},
  number = {5},
  journal = {Int. J. High Perform. Comput. Appl.},
  author = {Pazner,  Will and Kolev,  Tzanio and Camier,  Jean-Sylvain},
  year = {2023},
  pages = {578–599}
}

@article{Pazner24matrixfree,
  title = {Matrix-Free High-Performance Saddle-Point Solvers for High-Order Problems in {H}(div)},
  volume = {46},
  DOI = {10.1137/23m1568806},
  number = {3},
  journal = {SIAM J. Sci. Comput.},
  author = {Pazner,  Will and Kolev,  Tzanio and Vassilevski,  Panayot S.},
  year = {2024},
  pages = {B179–B204}
}

@article{Pazner25subspace,
  title = {Subspace and auxiliary space preconditioners for high-order interior penalty discretizations in {H}(div)},
  volume = {59},
  DOI = {10.1051/m2an/2025045},
  number = {4},
  journal = {ESAIM: Mathematical Modelling and Numerical Analysis},
  author = {Pazner,  Will},
  year = {2025},
  pages = {1909--1936}
}

@article{Phillips25,
  title = {Optimal Polynomial Smoothers and One‐Sided {V}‐Cycles for {P}oisson Problems},
  volume = {32},
  DOI = {10.1002/nla.70030},
  number = {4},
  journal = {Numerical Linear Algebra with Applications},
  author = {Phillips,  Malachi and Fischer,  Paul},
  year = {2025},
  pages = {e70030}
}

@book{Pope00turbulent,
  title = {Turbulent Flows},
  author = {Pope, Stephen B.},
  publisher = {Cambridge University Press},
  doi = {10.1017/cbo9780511840531},
  year = 2000,
  address = {Cambridge}
}

@article{Siklosi26,
  title = {Reduced and mixed precision turbulent flow simulations using explicit finite difference schemes},
  volume = {175},
  DOI = {10.1016/j.future.2025.108111},
  journal = {Future Generation Computer Systems},
  publisher = {Elsevier BV},
  author = {Siklósi,  Bálint and Sharma,  Pushpender K. and Lusher,  David J. and Reguly,  István Z. and Sandham,  Neil D.},
  year = {2026},
  pages = {108111}
}

@article{Still26discontinuous,
  title = {A discontinuous {G}alerkin consistent splitting method for the incompressible {N}avier--{S}tokes equations},
  volume = {458},
  DOI = {10.1016/j.cma.2026.119008},
  journal = {Comput. Meth. Appl. Mech. Engrg.},
  publisher = {Elsevier BV},
  author = {Still,  Dominik T. and Nebulishvili,  Natalia and Schussnig,  Richard and Kormann,  Katharina and Kronbichler,  Martin},
  year = {2026},
  pages = {119008}
}

@article{Stiller17,
  title = {Nonuniformly Weighted {S}chwarz Smoothers for Spectral Element Multigrid},
  volume = {72},
  DOI = {10.1007/s10915-016-0345-z},
  number = {1},
  journal = {Journal of Scientific Computing},
  publisher = {Springer Science and Business Media LLC},
  author = {Stiller,  J\"{o}rg},
  year = {2016},
  pages = {81--96}
}

@article{Sundar15,
  title = {Comparison of multigrid algorithms for high‐order continuous finite element discretizations},
  volume = {22},
  DOI = {10.1002/nla.1979},
  number = {4},
  journal = {Numerical Linear Algebra with Applications},
  publisher = {Wiley},
  author = {Sundar,  Hari and Stadler,  Georg and Biros,  George},
  year = {2015},
  pages = {664--680}
}

@inproceedings{Sundriyal26customized,
  series = {PASC '26},
  title = {Customized precision for discontinuous {G}alerkin methods using adaptive spectral block floating point},
  DOI = {10.1145/3815572.3815758},
  booktitle = {Proceedings of the Platform for Advanced Scientific Computing Conference},
  publisher = {ACM},
  author = {Sundriyal,  Shivam and B\"{u}ttner,  Markus and Kenter,  Tobias and Aizinger,  Vadym},
  year = {2026},
  pages = {1--12},
  collection = {PASC '26},
  address = {Bern}
}

@BOOK{Trottenberg01multigrid,
  Author = "Trottenberg, Ulrich and Oosterlee, Cornelius W. and Sch{\"u}ller, Anton",
  Title = "Multigrid",
  Publisher = "Elsevier Academic Press",
  Address = "London",
  Year = "2001"
}

@misc{Wells26wing,
  author = {Wells,  David and Knepley,  Matthew G. and Griffith,  Boyce E.},
  title = {{WING}: A simple windowed nonorthogonalized initial guess procedure for repeated matrix solves},
  eprint={2606.10132},
  archivePrefix={arXiv},
  primaryClass={math.NA},
  year = {2026},
  copyright = {Creative Commons Attribution Non Commercial No Derivatives 4.0 International}
}

@mastersthesis{wik22,
author = {Wik, Niklas},
title = {High-performance implementation of {H}(div)-conforming
elements for incompressible flows},
school = {Uppsala University},
year = {2022},
url = {https://uu.diva-portal.org/smash/get/diva2:1676116/FULLTEXT01.pdf}
}

@article{Witherden20impact,
  title = {Impact of Number Representation for High-Order Implicit Large-Eddy Simulations},
  volume = {58},
  DOI = {10.2514/1.j058434},
  number = {1},
  journal = {AIAA Journal},
  publisher = {American Institute of Aeronautics and Astronautics (AIAA)},
  author = {Witherden,  Freddie D. and Jameson,  Antony},
  year = {2020},
  pages = {184--197}
}

@article{SwirydowiczHighorder,
author = {Swirydowicz, Kasia and Chalmers, Noel and Karakus, Ali and Warburton, Tim},
year = {2017},
pages = {735--757},
title = {Acceleration of tensor-product operations for high-order finite element methods},
volume = {33},
journal = {Int. J. High Perform. Comput. Appl.},
doi = {10.1177/1094342018816368}
}

@misc{nvidiaHopperTuning,
  author       = {{NVIDIA Corporation}},
  title        = {{NVIDIA} {H}opper Tuning Guide},
  howpublished = {\url{https://docs.nvidia.com/cuda/hopper-tuning-guide}},
  year         = {2026},
  note         = {Accessed: May 2026}
}

@article{EichstaedtGPU,
author = {Eichstaedt, Jan and Peir\'o, Joaquim and Moxey, David},
year = {2022},
pages = {108624},
title = {Efficient vectorised kernels for unstructured high-order finite element fluid solvers on {GPU} architectures in two dimensions},
volume = {284},
journal = {Computer Physics Communications},
doi = {10.1016/j.cpc.2022.108624}
}

@article{kokkos,
  author={Trott, Christian R. and Lebrun-Grandié, Damien and Arndt, Daniel and Ciesko, Jan and Dang, Vinh and Ellingwood, Nathan and Gayatri, Rahulkumar and Harvey, Evan and Hollman, Daisy S. and Ibanez, Dan and Liber, Nevin and Madsen, Jonathan and Miles, Jeff and Poliakoff, David and Powell, Amy and Rajamanickam, Sivasankaran and Simberg, Mikael and Sunderland, Dan and Turcksin, Bruno and Wilke, Jeremiah},
  journal={IEEE Transactions on Parallel and Distributed Systems},
  title={Kokkos 3: Programming Model Extensions for the Exascale Era},
  year={2022},
  volume={33},
  number={4},
  pages={805-817},
  doi={10.1109/TPDS.2021.3097283}
}

@misc{amd_matrix_calculator,
  author       = {{Advanced Micro Devices, Inc.}},
  title        = {{AMD Matrix Instruction Calculator}},
  howpublished = {\url{https://github.com/ROCm/amd_matrix_instruction_calculator}},
  year         = {2026},
  note         = {Accessed: May 2026}
}

@misc{nvidia_mma_instruction,
  author       = {{NVIDIA Corporation}},
  title        = {Parallel Thread Execution ISA Version 9.3},
  howpublished = {\url{https://docs.nvidia.com/cuda/parallel-thread-execution/\#warp-level-matrix-instructions-for-mma}},
  year         = {2026},
  note         = {Accessed: May 2026}
}

@article{ADAMSChebyshevSmoother,
title = {Parallel multigrid smoothing: polynomial versus {G}auss--{S}eidel},
journal = {Journal of Computational Physics},
volume = {188},
number = {2},
pages = {593-610},
year = {2003},
issn = {0021-9991},
doi = {https://doi.org/10.1016/S0021-9991(03)00194-3},
url = {https://www.sciencedirect.com/science/article/pii/S0021999103001943},
author = {Mark Adams and Marian Brezina and Jonathan Hu and Ray Tuminaro}
}

@inproceedings{KronbichlerAllalen2018,
author="Kronbichler, Martin
and Allalen, Momme",
editor="Bungartz, Hans-Joachim
and Kranzlm{\"u}ller, Dieter
and Weinberg, Volker
and Weism{\"u}ller, Jens
and Wohlgemuth, Volker",
title="Efficient high-order discontinuous {G}alerkin finite elements with matrix-free implementations",
booktitle="Advances and New Trends in Environmental Informatics",
year="2018",
publisher="Springer International Publishing",
address="Cham",
pages="89--110",
isbn="978-3-319-99654-7"
}

@article{Cui2025PatchSmootherGPU,
	author = {Cui, Cu and Grosse-Bley, Paul and Kanschat, Guido and Strzodka, Robert},
	title = {An Implementation of Tensor Product Patch Smoothers on {GPUs}},
	journal = {SIAM Journal on Scientific Computing},
	volume = {47},
	number = {2},
	pages = {B280-B307},
	year = {2025},
	doi = {10.1137/24M1642706},
	URL = {https://doi.org/10.1137/24M1642706}
}

@inproceedings{Rudi2015EarthMantle,
	author = {Rudi, Johann and Malossi, A. Cristiano I. and Isaac, Tobin and Stadler, Georg and Gurnis, Michael and Staar, Peter W. J. and Ineichen, Yves and Bekas, Costas and Curioni, Alessandro and Ghattas, Omar},
	title = {An extreme-scale implicit solver for complex {PDEs}: highly heterogeneous flow in earth's mantle},
	year = {2015},
	isbn = {9781450337236},
	publisher = {Association for Computing Machinery},
	address = {New York, NY, USA},
	url = {https://doi.org/10.1145/2807591.2807675},
	doi = {10.1145/2807591.2807675},
	booktitle = {Proceedings of the International Conference for High Performance Computing, Networking, Storage and Analysis},
	articleno = {5},
	numpages = {12},
	location = {Austin, Texas},
	series = {SC '15}
}

@InProceedings{Bauer2020TerraNeo,
	author="Bauer, Simon
	and Bunge, Hans-Peter
	and Drzisga, Daniel
	and Ghelichkhan, Siavash
	and Huber, Markus
	and Kohl, Nils
	and Mohr, Marcus
	and R{\"u}de, Ulrich
	and Th{\"o}nnes, Dominik
	and Wohlmuth, Barbara",
	editor="Bungartz, Hans-Joachim
	and Reiz, Severin
	and Uekermann, Benjamin
	and Neumann, Philipp
	and Nagel, Wolfgang E.",
	title="TerraNeo---Mantle Convection Beyond a Trillion Degrees of Freedom",
	booktitle="Software for Exascale Computing - SPPEXA 2016-2019",
	year="2020",
	publisher="Springer International Publishing",
	address="Cham",
	pages="569--610",
	isbn="978-3-030-47956-5"
}

@misc{Vacek2026Mixed,
	title={Mixed precision multigrid with smoothing based on incomplete Cholesky factorization}, 
	author={Petr Vacek and Hartwig Anzt and Erin Carson and Nils Kohl and Ulrich Rüde and Yu-Hsiang Tsai},
	year={2026},
	eprint={2511.04566v2},
	archivePrefix={arXiv},
	primaryClass={math.NA},
}
    }

\end{document}